\documentclass[prd,a4paper,superscriptaddress,nofootinbib,10pt]{revtex4} 
\usepackage{graphicx}
\usepackage{amssymb}
\usepackage{amsmath}
\usepackage{lscape}
\usepackage{mathrsfs}
\usepackage{textcomp}
\usepackage{epsfig}
\usepackage{slashed}
\usepackage{comment}
\usepackage{enumerate}
\renewcommand{\d}{\mathrm{d}}
\begin{document}

\title{Observables in Quantum Mechanics and the Importance of Self-adjointness}

\author{Tajron Juri\'c}
\email{tjuric@irb.hr}
\affiliation{Rudjer Bo\v{s}kovi\'c Institute, Bijeni\v cka  c.54, HR-10002 Zagreb, Croatia}

\date{\today}

\begin{abstract}

We are focused on the idea that observables in quantum physics are a bit more then just hermitian operators and that this is, in general, a ``tricky business''.
The origin of this idea comes from the fact that there is a subtle difference between symmetric, hermitian, and self-adjoint operators which are of
immense importance in formulating Quantum Mechanics. The theory of self-adjoint extensions is
presented through several physical examples and some emphasis is given on the physical implications and applications.

\end{abstract}

\maketitle

\section{Introduction}

Proper quantization of physical systems requires a correct  definition of physical observables (such as the Hamiltonian, momentum, etc.) as self-adjoint operators in an appropriate Hilbert space and their spectral analysis. This problem is a much more subtle issue then the mere prescription of assigning a hermitian operator (or matrix) to a classical observable. Solution to this problem is not straightforward and goes beyond the finite-dimensional hermiticity condition when dealing with nontrivial quantum systems such as systems on  manifolds with boundaries or with singular interactions. These nontrivial quantum systems with singular potentials, both relativistic and non-relativistic, with or without boundaries, play a very important role in physics, and a consistent and well defined quantization needs a considerable amount of techniques and results usually  found only in different advanced chapters on functional analysis. However, these advanced functional analytical methods usually go beyond the scope of the mathematical apparatus usually presented in standard textbooks on Quantum Mechanics (QM) for physicists\footnote{There are some exceptions such as \cite{iznimke} which are mainly intended for mathematically minded physicists and mathematicians.}, e.g., \cite{standard}, and unfortunately this remains unchanged even in more modern textbooks \cite{nove}. The aim of this paper is twofold. The first goal is of more pedagogical nature, namely, to convince the reader-physicist that one must take a lot of caution  when reading standard textbooks on QM for physicists and ``blindly'' applying the notions and prescriptions from such textbooks to nontrivial quantum systems. The second goal is to show the natural appearance (and necessity) of anomalies in QM and to review some of the recent (and not so recent) applications and bring the reader to the frontier of some of the latest research where the self-adjoint issues play a crucial role.\\

The mathematical framework of QM is functional analysis, more precisely, the theory of linear operators in Hilbert spaces. This is a quite ``subtle science'' and  to completely master it would take a considerable amount of time. It is for this reason that standard textbooks on QM for physicists usually just present a rather simplified version of the relevant parts of functional analysis in the  form of brief ``rules'' (sometimes even called axioms\footnote{See  Appendix A for the usual postulates of QM.}) which are then applied to ``well behaved systems'' (such as harmonic oscillator, particle in a box,  etc.) where many of the mathematical subtleties are then necessarily left aside or ``swept under the rug''. These ``rules'' are, more or less, based on our experience in finite-dimensional linear algebra and dealing with algebra of matrices, which can often be misleading and lead to some paradoxes when taken too literally. We will outline some of these paradoxes in Section II in order to motivate the reader-physicist.\\ 

Observables are usually presented as hermitian operators (matrices). Hermitian matrices have important properties like real eigenvalues, the corresponding eigenvectors are orthogonal and span the whole finite-dimensional Hilbert space, etc. However, all these properties are not ensured by the hermiticity condition in general infinite-dimensional Hilbert space. Hermiticity condition is often just replaced with the symmetricity condition, which only ensures for the expectation values of observables to be real \cite{Cintio}, while the rest of the properties can only be grasped with imposing a more subtle condition called self-adjoitness.\\

A crucial subtlety is that an unbounded\footnote{See Appendix B for the definition of bounded operators, Appendix C for the details on Hilbert space, Appendix D for the definition of the adjoint operator and Appendix E for the definition and properties of a SA operator.} self-adjoint (SA) operator cannot be defined in the whole Hilbert space, i.e. it can not act on an arbitrary QM state\footnote{Here ``state'' is used in the physical jargon and actually means an element of the Hilbert space $\mathcal{H}$ (or an equivalence class of elements up to a phase). The state of a system is a positive linear map $\rho:\mathcal{H}\longrightarrow\mathcal{H}$ for which $Tr(\rho)=1$. States can be pure or mixed. A state $\rho_\psi$ is called pure if it maps $\psi\longmapsto\frac{(\psi,\ \cdot\ )}{(\psi,\psi)}, \ \forall \psi\in\mathcal{H}$. Thus, we can associate to each pure state $\rho_\psi$ an element in $\mathcal{H}$. However, this correspondence
is not one-to-one, and one should have this fact in mind. The physicist refer to the operators $\rho$ as density matrices and often in Dirac notation just write $\rho=\left|\psi\right\rangle\left\langle \psi\right|$.}, which is often assumed in most textbooks when presenting  operator or canonical quantization. One must be aware that  an operator without its domain of definition is not well defined. An operator is not only a rule of acting, but also comes with a domain in a Hilbert space to which this rule is applicable (see \cite{Cintio} for an elaborate discussion). In the case of unbounded operators, the same rule of acting for different domains might generate different operators with sometimes completely different properties. The pivotal example comes from examining a differential equation $f^{''}=\lambda f$ for different boundary conditions. The differential equation provides the rule of acting while boundary conditions selects the domain of functions, and depending on the domain chosen, different values for the parameter $\lambda$ are allowed leading to different physical interpretations or regimes.  Having this in mind, it is clear that  when provided with a rule of acting, it is  only the appropriate choice of a domain for a QM observable that makes it a SA operator. The main problems and unwanted paradoxes are related to this point. But only when we understand where and why these problems occur, can we look to find ways to resolve them.\\

The theory of SA extensions of unbounded symmetric operators provides the main tool for dealing with these problems. The theory of SA extensions provides us with all possible, or better say, allowable domains for a symmetric operator which render it SA. In general, these extensions are not unique, if they are possible at all. From the physicist point of view, this implies that upon performing the quantization of a nontrivial system, we are generally presented with various non-equivalent possibilities for its quantum description. The general theory of SA extensions provides us with all that mathematics can offer to a physicist. Of course, the physical interpretation of the available SA extensions is a purely physical problem. Any extension is a certain prescription for the behavior of a physical system under consideration near its boundaries or singularities. In a sense,  different extensions classify the types of quantum-mechanically allowable interactions, and is up to the physicist to investigate all of them and chose the one which fits the experimental results.  It is also believed that each extension can be understood through an appropriate regularization and a subsequent limit process, although this is a totally new problem that stands on its own. \\

Self-adjointness and the theory of SA extensions are known to play important roles in a variety of physical contexts, including systems with a confined particle \cite{confined1, confined2, confined3}, Aharonov-Bohm effect \cite{AB, Salem:2019xgk}, graphene \cite{graph}, two and three dimensional delta function potentials \cite{delta}, heavy atoms \cite{heavy}, singular potentials \cite{singularp}, Calogero models \cite{calogero}, anyons \cite{anyons}, anomalies \cite{anomalies, an1}, $\zeta$-function renormalization \cite{renorm}, scattering theory \cite{botelho}, particle statistics \cite{stat}, black holes \cite{bh}, integrable system \cite{im}, Klein-Gordon equation \cite{amuch}, renormalons in QM \cite{renormalon}, quasinormal modes \cite{qnm}, supersymmetric QM \cite{sqm} and toy models for strings \cite{Fredenhagen:2003ut}, spectral triple \cite{connes}, noncommutative field theories \cite{ncft1, ncft2, ncft3}, resolving the spacetime singularities \cite{singularity} and even plays a very important role in classical physics since it underlies the conservation of entropy of Hamiltonian systems \cite{bondar}.\\

The outline of the paper is as follows. In section II we motivate the reader by giving some paradoxes of conventional QM, and together with the Appendices A, B, C, D, and E we establish the notation and give an overview on the formalism of  QM, Hilbert spaces and properties of certain operators. In section III we give a pedagogical, step by step introduction to the  subtleties of SA operators and its extensions. This is done by going through simple examples.  In section IV we discuss the existence of new bound states, anomalies and give a simple proof of Pauli's theorem. In section V we comment on the canonical quantization in higher dimensions and finally in section VI we finish with some remarks. The more rigorous\footnote{Claims like ``rigorous definition'' or ``rigorous proof'' are also very often used in the physicist jargon. The thing is that objects are either well defined or not, or the claim is proven or not. There is no degree of ``how much'' something is or can be proven, but when physicist says that something is rigorously done, it just means that ``all the necessary''  assumptions that deal with questions of convergence, completeness, domains etc. are assumed to be valid (or better say omitted). Or simply, the nonrigouros proofs or definitions seems to work well in several cases and one just takes it for granted that the results apply for other situations also.} definitions and some discussions are given in the Appendices B, C, D, E and F at the end of the paper.\\

\section{Motivation}
In order to motivate a physicist to learn about the subtleties of functional analysis, we will present some paradoxes that even appear in simple QM systems\footnote{As they are usually presented in conventional undergraduate or graduate physics courses.} obtained in the idealized scheme of operator canonical quantization\footnote{See Appendix A for postulates of QM}. We will see that if one follows these postulates literally, one can arrive at certain paradoxes, i.e.  contradictions with the well-known statements.\\

For this purpose, let us first consider a system that consists of a free non-relativistic particle of mass $m$ moving on a real interval $(a,b)$. Depending on the values of $a$ and $b$, the interval can be finite, infinite or semi-finite. Also we can include or not include the boundary, but that we will specify later. In classical mechanics, this system is modeled by the phase space $\mathcal{P}=T^{*}M=(a,b)\times\mathbb{R}$, that is a strip  given by the range of the position $x\in(a,b)$ and momentum $p\in\mathbb{R}$ of the particle. The Poisson bracket of the position $x$ and momentum $p$ is $\left\{x,p\right\}=1$ and the time evolution is governed by the Hamiltonian equations. The Hamiltonian for free motion is given by $H=p^2/2m$. Depending on the values of $a$ and $b$, that is, if one of them is finite, we are dealing with a phase space which is a manifold with boundaries. In such situation, the behavior of the particle near that boundary must be specified by some extra conditions such as elastic/nonelastic/plastic reflection, delay, trapping or something else. This is more or less all the information we need to fully characterize our classical system and now we wish to quantize it.\\

In order to quantize our system we have to use the postulates of QM as they are presented in  Appendix A. In doing so, we may notice that it seams that we don't encounter problems with possible  boundaries of the classical phase space. Namely, the canonical observables for the QM system are represented by SA operators  $\hat{x}$ for the position and $\hat{p}$ for the momentum of the particle, which satisfy the 
canonical commutation relations
\begin{equation}\label{cancom}
[\hat{x},\hat{p}]=i\hbar\left\{x,p\right\}=i\hbar.
\end{equation}
In the so called $x$-representation of \eqref{cancom}  the Hilbert space $\mathcal{H}$ of all possible quantum states is actually the space of functions $\psi(x)$ that are square-integrable\footnote{Here one should be aware of the full details of the construction of the Hilbert space $L^2$. Namely, one needs first a measurable space  $(M,\Sigma, \mu)$ (here $M$ is a set, $\Sigma$ is the Borel-$\sigma$-algebra and $\mu$ is a measure)  together with an equivalence relation 
$f\sim g \Leftrightarrow \left\|f\right\|_2=\left\|g\right\|_2$, so that 
$$L^2(M):=\mathcal{L}^2 / \sim =\left\{[f] | f\in \mathcal{L}^2 \right\}$$ where $$\mathcal{L}^2(M,\Sigma, \mu):=\left\{f:M\rightarrow\mathbb{C}|\ \Re(f)\ \text{and}\ \Im(f)\ \text{are measurable and}\ \int_{M}\left|f\right|^2d\mu<\infty\right\} $$ with the inner product $(\cdot, \cdot): L^2 \times L^2 \rightarrow\mathbb{C}$ and $([f],[g])\mapsto\int_{M}\bar{f}gd\mu$.} on the interval $(a,b)$. This Hilbert space is often denoted by $\mathcal{H}=L^2(a,b)$. In this representation, the position operator $\hat{x}$ is the operator of multiplication by $x$, namely
\begin{equation}
\hat{x}\psi(x)=x\psi(x)
\end{equation}
and its spectrum\footnote{In finite dimensional space the spectrum of an operator is the set of its eigenvalues. In general Hilbert space the spectrum of a SA operator is much more. It can have the point spectrum part (related to the set of eigenvalues) and the continuous spectrum part (usually related to generalized eigenfunctions that physicist sometimes call scattering states)\cite{sae}.} is given by $\sigma(\hat{x})=(a,b)$. The momentum operator $\hat{p}$ can be uniquely represented using the differentiation operator $\frac{\d}{\d x}$ via
\begin{equation}
\hat{p}=-i\hbar\frac{\d}{\d x}, \quad \hat{p}\psi(x)=-i\hbar\psi^\prime (x).
\end{equation}
Clearly, the canonical commutation relation \eqref{cancom} holds. Other quantum observables can be constructed  from the classical observable $f(x,p)$ using the correspondence principle and  in general they are represented by certain differential operators
\begin{equation}
\hat{f}=f(x,-i\hbar\frac{\d}{\d x})+O(\hbar).
\end{equation}
In particular the free quantum Hamiltonian is given by
\begin{equation}\label{Ham}
\hat{H}=\frac{\hat{p}^2}{2m}=-\frac{\hbar^2}{2m}\frac{\d^2}{\d x^2}.
\end{equation}

So far everything seams well defined and unambiguous, since in case the boundaries of the interval are finite $\left|a\right|,\left|b\right|<\infty$ one can consider the Hilbert space $L^2(a,b)$ as a subspace of $L^2(\mathbb{R})$ where all the states vanish outside the interval $(a,b)$ (they are compactly supported functions), while the observables $\hat{x}$ and $\hat{p}$ are restrictions of the well-known SA operators defined on $L^2(\mathbb{R})$ to this subspace. Notice that for the finite interval $[a,b]$, the position operator $\hat{x}$ becomes a bounded SA operator defined everywhere \cite{r-s-1}. Considering the momentum $\hat{p}$ and Hamiltonian $\hat{H}$ also as SA operators satisfying\footnote{This is actually just the symmetric condition, which in the final-dimensional case is hermiticity.}
\begin{equation}\label{herm}
(\varphi, A\psi)=(A\varphi, \psi), \quad A\in\left\{\hat{x}, \hat{p}, \hat{H} \right\}, \quad \forall \varphi, \psi\in\mathcal{H}
\end{equation}
and the commutation relations
\begin{equation}
[\hat{x},\hat{p}]=i\hbar, \quad [\hat{x},\hat{H}]=0
\end{equation}
 finishes the canonical quantization procedure.\\

In what follows we will apply the above procedure to certain situations and soon realize that it might lead to certain inconsistencies, i.e. paradoxes.\\

\subsection{Paradox 1}
Let $\psi_{p}$ be a normalized eigenvector of the SA operator, $\hat{p}\psi_{p}=p\psi_{p}$.  Since the operators $\hat{p}$ and $\hat{x}$ are SA, i.e. satisfy \eqref{herm} , we can calculate the  following
\begin{equation}\begin{split}
(\psi_{p},[\hat{x},\hat{p}]\psi_{p})&=(\psi_{p},\hat{x}\hat{p}\psi_{p})-(\psi_{p},\hat{p}\hat{x}\psi_{p})\\
&=p(\psi_{p},\hat{x}\psi_{p})-(\hat{p}\psi_{p},\hat{x}\psi_{p})\\
&=p\left[(\psi_{p},\hat{x}\psi_{p})-(\psi_{p},\hat{x}\psi_{p})\right]=0,
\end{split}\end{equation}
where we used the definition of the commutator $[A,B]=AB-BA$ and the self-adjointness of $\hat{p}$.  On the other hand, if we first use  \eqref{cancom} we get
\begin{equation}
(\psi_{p},[\hat{x},\hat{p}]\psi_{p})=i\hbar(\psi_{p},\psi_{p})=i\hbar\neq 0,
\end{equation}
arriving at a contradiction.

In addition, the canonical  commutation relations, due to Cauchy-Schwarz-Bunyakovski inequality, imply the  famous Heisenberg uncertainty relation
\begin{equation}\label{hup}
\Delta \hat{x} \Delta \hat{p} \geq\frac{\hbar}{2},
\end{equation}
where $\Delta \hat{x}=\sqrt{\left\langle \hat{x}^2\right\rangle_{\psi}-\left\langle \hat{x}\right\rangle^{2}_{\psi}}$ and $\Delta \hat{p}=\sqrt{\left\langle \hat{p}^2\right\rangle_{\psi}-\left\langle \hat{p}\right\rangle^{2}_{\psi}}$ are the dispersions of the position and momentum for any state $\psi$ of a particle. However,  when dealing with the finite interval $[a,b]$ and for $\psi=\psi_{p}$ (the eigenvector of momentum), we get $\Delta \hat{x}\leq b-a$, $\Delta \hat{p}=0$, which leads to $\Delta \hat{x} \Delta \hat{p} =0$, and that is in a contradiction with \eqref{hup}.\\

The resolution of these apparent paradoxes is different for different types of interval.  Namely, depending on the type of the interval (finite, semi-finite or infinite), either a SA momentum operator does not exist, or it exists but has no eigenvectors, or even if such vectors exist, they do not belong to the domain of the operator $\hat{p}\hat{x}$. Therefore, in order to encounter aforementioned paradoxes, the knowledge about the domain of the operators is crucial and unavoidable in practical calculations. In addition, one can easily see that for the case of a semi-finite or  finite interval, the canonical commutation relation \eqref{cancom} and the uncertainty principle \eqref{hup} are in contradiction. Actually the origin of the problem is that we are dealing with unbounded operators and the very concept of calculating commutation relations for unbounded operators is not well defined \cite{Cintio}.

\subsection{Paradox 2}
Observe that the canonical commutation relation \eqref{cancom} cannot be realized for observables $\hat{x}$ and $\hat{p}$ acting as operators (matrices) on a non-trivial finite-dimensional Hilbert space $\mathcal{H}$ if the Planck's constant $\hbar$ is different from zero \cite{Zeidler}. To see this let us assume that $\hat{x}$ and $\hat{p}$ are SA linear operators $\hat{x},\hat{p}:\mathcal{H}\rightarrow\mathcal{H}$ such that \eqref{cancom} holds. Now we take the trace of \eqref{cancom} and calculate
\begin{equation}\begin{split}
\mathrm{Tr}([\hat{x},\hat{p}])&=\mathrm{Tr}(i\hbar I)  \\
\mathrm{Tr}(\hat{x}\hat{p})-\mathrm{Tr}(\hat{p}\hat{x})&=i\hbar \mathrm{Tr}(I) \\
0&=i\hbar\times dim(\mathcal{H}) ,
\end{split}\end{equation}
where we used the cyclicitiy of the trace $\mathrm{Tr}(AB)=\mathrm{Tr}(BA)$, $I$ is the identity operator and $dim(\mathcal{H})$ is the dimension of the Hilbert space. The conclusion is that either $\hbar=0$ or we have to abandon the idea of having a finite-dimensional Hilbert space and accept the fact that at least one of the operators in question is unbounded and that the issue of trace-classness emerges too.

\subsection{Paradox 3}
Let us now consider a free particle moving on a finite interval $[0,a]$. We can treat this motion as the one governed by the Hamiltonian \eqref{Ham} containing  an infinite rectangular potential well. In doing so, we get  that the eigenvalues of the Hamiltonian and corresponding eigenfunctions are (as can be found in any  QM textbook):
\begin{equation}
\hat{H}\psi_{n}=E_{n}\psi_{n}, \quad E_{n}=\frac{\hbar^2\pi^2}{2ma^2}n^2, \quad \psi_{n}(x)=\sqrt{\frac{2}{a}}\ \text{sin}\left(\frac{n\pi}{a}x\right), \quad n\in\mathbb{N}.
\end{equation}
Note that the set\footnote{$\left\{a_n\right\}^{\infty}_{1}$ denotes a set of members of the  sequence $a_n$ starting from $n=1$ to $n=\infty$.} $\left\{\psi_n\right\}^{\infty}_{1}$ of the eigenfunctions is indeed an orthonormal basis in $L^2[0,a]$, which confirms the self-adjointness of the Hamiltonian\footnote{This claim stands even rigorously, because if for a symmetric operator $H$ its domain $D(H)\subset\mathcal{H}$ contains the orthonormal basis of $\mathcal{H}$, then $H$ is SA (see \cite{Cintio}).}.\\

As we pointed out before, the claim that two commuting SA operators have common eigenvectors is often assumed by physicists. If this is so, and if the spectrum of the commuting SA operators are nondegenerate, then its eigenvectors must be eigenvectors of another SA operator. In the case of the free particle on a finite interval, the two commuting SA operators are the momentum $\hat{p}$ and  Hamiltonian $\hat{H}$, and since the spectrum of $\hat{H}$ is nondegenerate the eigenfunctions of $\hat{H}$ must be eigenfunctions of $\hat{p}$ also. But we can check that by calculating
\begin{equation}
\hat{p}\psi_{n}=-i\hbar\sqrt{\frac{2}{a}}\frac{n\pi}{a}\ \text{cos}\left(\frac{n\pi}{a}x\right)\neq p_{n}\psi_{n}
\end{equation}
for any $n$, and immediately see that it contradicts the above assertion.\\

This paradox is again a consequence of the ill defined assumption that $\hat{p}$ and $\hat{H}$ commute. Each of these operators has its own domain, and the commutator can not be defined in general. In particular, also the  na\"{i}ve belief that the Hamiltonian can be represented just as $\hat{H}=\hat{p}^2/2m$, that is, as a simple composition of the operator $\hat{p}$ without referring to both of its domains, could lead to this type of paradoxes.

\subsection{Paradox 4}
Observables are linear SA operators in Hilbert space. When the Hilbert space in question is finite, these operators are Hermitian matrices, but when dealing with square integrable functions as our quantum states then we are dealing with infinite-dimensional Hilbert spaces and the analog of matrix elements of (unbounded) operators are not well defined. However, in most standard QM textbooks it is claimed that the matrix elements $A_{mn}=(e_{m},\hat{A}e_{n})$ of an operator $\hat{A}$ with respect to an orthonormal basis $\left\{e_{n}\right\}^{\infty}_{1}$ completely determine the operator $\hat{A}$. This is justified by the following reasoning
\begin{equation}\begin{split}
\forall\psi\in{\mathcal{H}}\Longrightarrow\psi&=\sum^{\infty}_{n=1}\psi_{n}e_{n}, \quad \psi_{n}=(e_{n},\psi), \quad \hat{A}e_{n}=\sum^{\infty}_{m=1}A_{mn}e_{m},\\
\hat{A}\psi&=\sum^{\infty}_{n=1}\psi_{n}\hat{A}e_{n}=\sum^{\infty}_{m=1}\left(\sum^{\infty}_{n=1}A_{mn}\psi_{n}\right)e_{m}.
\end{split}\end{equation}
Then the adjoint $\hat{A}^\dagger$ of $\hat{A}$ could be defined as an operator whose matrix elements are given by
\begin{equation}
\left(\hat{A}^\dagger\right)_{mn}=\left(e_{m},\hat{A}^\dagger e_{n}\right)=\left(\hat{A}e_{m}, e_{n}\right)=\overline{\left(e_{n},\hat{A} e_{m}\right)}=\overline{A_{nm}}.
\end{equation}
In other words, a SA operator $\hat{A}^\dagger =\hat{A}$ is defined as an operator whose matrix is hermitian, i.e. $A_{mn}=\overline{A_{nm}}$.\\

If the above is true, let us apply it to some simple example. To this matter we can look at a matrix $p_{nm}=\left(e_{n},\hat{p} e_{m}\right)$ that corresponds to the momentum operator in the Hilbert space $L^2[0,l]$ with respect to some orthonormal basis $\left\{e_{n}\right\}^{\infty}_{0}$. Let us use a particular basis defined by
\begin{equation}\label{orto}
e_{n}(x)=\sqrt{\frac{2}{l}}\ \text{cos}\left(\frac{n\pi}{l}x\right), \quad \forall n\in\mathbb{N}_{+}.
\end{equation}
A straightforward check of the hermiticity condition for $p_{mn}$  yields
\begin{equation}
\overline{p_{nm}}=p_{mn}+\left[e_{m}(l)e_{n}(l)-e_{m}(0)e_{n}(0)\right]\neq p_{mn}, \quad m+n=2k+1,
\end{equation}
where we used partial integration. It is obvius that the matrix $p_{mn}$ is not hermitian, leading to jet another paradox. This paradox is related to the fact that the orthonormal basis \eqref{orto} does not belong to the domain of any SA operator $\hat{p}$ from the whole family of admissible momentum operators.\\

\subsection{Paradox 5}
Let us consider a particle confined on a circle. The obvious Hilbert space is given by the square integrable periodic functions\footnote{Actually, this just a dense subspace and its completion is the Hilbert space $L^2[0,2\pi]$.}, i.e.
\begin{equation}\label{circH}
\mathcal{H}=\left\{\psi\ | \ \psi(\theta+2\pi)=\psi(\theta);\ \int^{2\pi}_{0}d\theta \left|\psi\right|^2<\infty\right\}.
\end{equation}
The position operator of the particle is given by the multiplication by the angle coordinate
\begin{equation}
\hat{x}\psi(\theta)=\theta\psi(\theta),
\end{equation}
while the momentum operator is the multiplicative of the derivative with respect to $\theta$
\begin{equation}
\hat{p}\psi(\theta)=-i\hbar \frac{\partial \psi}{\partial\theta}
\end{equation}
in order to satisfy \eqref{cancom}. With such defined $\hat{x}$ and $\hat{p}$ it is easy to see that they are SA (in the sense that they satisfy \eqref{herm}). The normalized eigenvectors of the momentum operator $\hat{p}$ and its spectrum are given by
\begin{equation}
\psi_{n}=\frac{\text{e}^{in\theta}}{\sqrt{2}}, \quad \sigma(\hat{p})=n\pi, \ \ n\in\mathbb{Z}.
\end{equation}
The $\left\{\psi_n\right\}^{\infty}_{-\infty}$ form a complete orthonormal basis for the Hilbert space $L^2[0,2\pi]$. Notice that $\hat{x}$ is a bounded operator since $\theta\in\left[0,2\pi\right\rangle$, thereby when calculating the uncertainty $\Delta\hat{x}$ in the state $\psi_n$ one gets a finite value, while for the corresponding momentum uncertainty we get $\Delta \hat{p}=0$, again a contradiction with \eqref{cancom}. The reason for this paradox is that the states $\psi_n$ are not in the domain of the $\hat{p}\hat{x}$ operator. Namely, when calculating $\hat{p}\hat{x}\psi_n=\hat{p}(\hat{x}\psi_n)$ one sees that the state $\hat{x}\psi_n$ is not periodic and not an element of $\mathcal{H}$, let alone in the domain of $\hat{p}$, rendering the canonical commutation relation \eqref{cancom}  ill defined.\\

The number of these type of paradoxes can be extended \cite{sae, example, Cintio, Gieres:1999zv}.

\section{Basic ideas and results on SA operators}
We have seen that it is crucial to think of an operator as acting on some domain. The domain is encoded in the boundary conditions. For better understanding this issue   we shall derive the basic definitions and results, which are going to be illustrated through very simple examples, step by step. The purpose of these section is to illustrate the procedure of von Neumann in simple terms using various examples from physics. Here we have mostly used \cite{sae} from which we have freely borrowed.\\

In order to solve a differential equation, we need to impose some boundary conditions. We know that the solution of this differential equation highly depends on these imposed boundary conditions. Each boundary condition corresponds to some physically different situation, that is, nature of the interaction is encoded in the boundary condition. So, we can ask ourselves a natural question: ``What are the possible choices of boundary conditions for a given operator representing an observable in QM?''. Before we address this question we must remember that observables in QM must have real eigenvalues. These operators are called hermitian or SA, although as we shall see below, these two concepts are not quite the same. Having this in mind, we can thus ask the following question: ``What are the possible boundary conditions that can be imposed on an operator in QM such that it is SA?''. The answer to this question can be obtained from the pioneering work of von Neumann on SA extensions of operators in QM (see \cite{sae} and Appendix F).\\

\subsection{The momentum operator on a finite interval}
To order to obtain the spectrum\footnote{Physicist often thinks of the spectrum as the set of eigenvalues (point spectrum) which come from solving differential equations with certain boundary conditions and imposing square integrability of the solutions. However, to grasp the continuous part of the spectrum (or energy of the scattering states), physicist solves the same differential equation but ``drops'' the square-integrability condition and looks for the solution outside the Hilbert space. This at first glance strange procedure is imposed by the Dirac notation and its justification can be found in the theory of rigged Hilbert spaces (or Gelfand triples). For more details see Appendix G and \cite{Zeidler}\cite{Madrid}.} of an observables such as momentum or Hamiltonian, physicist often wants to determine it by solving some differential eigen-equation and for that reason one has to impose some suitable boundary conditions. For a generic observable we denote its corresponding operator by  $T$ that acts on some Hilbert space $\mathcal{H}$ (actually the completion of the space of square integrable functions that we often call wave functions). However, this operator $T$ is usually a differential operator and when trying to solve the corresponding eigen-equation one needs to impose some boundary conditions. The very existence of such boundary conditions implies that $T$ acts only on some subset\footnote{We also assume that $D(T)$ is dense in $\mathcal{H}$, that is $\overline{D(T)}=\mathcal{H}$, which means that the topological closure of $D(T)$ is the same as $\mathcal{H}$. $D(T)$ may not be a complete space, so dense means that the space together with limit vectors of all of its Cauchy sequences is exactly $\mathcal{H}$.} $D(T)\subset\mathcal{H}$, and not on all of $\mathcal{H}$. The subset $D(T)$ is called the domain of $T$. Elements of $D(T)$ are defined as elements of the Hilbert space that obey the imposed boundary condition and it is immediately clear that $D(T)$ is not the whole Hilbert space $\mathcal{H}$, since there will, in general, be various elements in $\mathcal{H}$ which may not obey the imposed boundary condition. The  domain $D(T)$ of $T$ is an integral part of the definition of the operator $T$, and one should think of an operator as the rule of acting $T$ together with its domain $D(T)$, because different domains can render different properties of the operator $T$ (like spectrum, boundlessness, self-adjointness, etc.).  To illustrate the aforementioned, let us consider the example of a momentum operator given by\footnote{For simplicity we take the unit mass and $\hbar=1$ system of units. }
\begin{equation}\label{p}
\hat{p}=-i\frac{\d}{\d x}
\end{equation}
on a finite interval, and for simplicity we choose $x\in[0,1]$. To define the domain of $\hat{p}$ we assume that the Hilbert space should consist of square integrable functions on the finite interval and that the boundary condition should reflect the physical intuition that the wave function describing the particle should vanish on the boundaries. Therefore we define the domain of $\hat{p}$ as\footnote{From now on we understand that all derivatives are in the weak sense.} 
\begin{equation}\label{D(p)}
D(\hat{p})=\left\{\psi\ \ |\ \psi(0)=\psi(1)=0,\ \psi\ \text{absolutely continuous and}\ \psi\in L^2[0,1]\right\},
\end{equation}
where\footnote{A function $f:[a,b]\longrightarrow\mathbb{C}$ is absolutely continuous if it has a derivative $f^\prime$ almost everywhere, the derivative is Lebesgue integrable and $$f(x)=f(a)+\int^{x}_{a}f^\prime (t)dt, \quad \forall x\in[a,b].$$} $L^2[0,1]$ is the Hilbert space of  (the completion) of square integrable functions on the finite interval $[0,1]$. In what follows, whenever speaking about the momentum operator $\hat{p}$ we have the domain \eqref{D(p)} in mind.\\

Apart from the domain of the operator, we are interested in some special properties. Sort of generalizations, or better to say  refinements, of the hermiticitiy condition in finite-dimensional Hilbert spaces. For this purpose we give few definitions that will be of much use later. The Hilbert space is equipped with a sesquilinear inner product\footnote{For more details see Appendix C.} and for any two elements $\xi,\eta\in\mathcal{H}$ it is denoted by $(\xi,\eta)$. An operator $T$ with a domain $D(T)\subset\mathcal{H}$ is said to be  \textsl{symmetric} if the relation\footnote{This reduces to the hermicity condition in finite-dimensional Hilbert spaces.}
\begin{equation}
(\xi,T\eta)=(T\xi,\eta)
\end{equation}
holds $\forall\xi,\eta\in D(T)$. Note that only this relation is often used to define SA operators in elementary QM, and the domain issues are left aside. Strictly speaking this condition only defines a symmetric operator and we shall soon see the difference between this relation and the condition for self-adjointness.\\

Now, let $T^\dagger$ denotes the operator adjoint of $T$. In order to fully specify the operator $T^\dagger$, we need to define its domain $D(T^\dagger)$ also. Since we are not interested in any operator\footnote{See Appendix D.}, but rather special ones (hermitian, symetric, self-adjoint) we will define the adjoint of a symmetric operator. For that, let as consider a symmetric operator $T$ and let $\xi\in D(T)$. The adjoint $T^\dagger$ satisfies \begin{equation}
T^\dagger\xi=T\xi
\end{equation}
and its domain  $D(T^\dagger)$ is defined as
\begin{equation}
D(T^\dagger)=\left\{\psi\in\mathcal{H}\ \ | \  \exists\eta=T^\dagger\xi\in\mathcal{H}\ \forall \xi\in D(T)\ \ \text{such that} \ \ (\psi, T\xi)=(\eta,\xi)\right\}
\end{equation}
It is important to note that $D(T^\dagger)$  is in general different from $D(T)$ and by construction we have
\begin{equation}
(T^\dagger \eta, \xi)=(\eta, T\xi)\ \text{and} \ \ (\xi,T^\dagger \eta)=(A\xi, \eta) \ \forall \xi\in D(T) \ \text{and}\ \eta\in D(T^\dagger)
\end{equation} Finally, an operator is called \textsl{self-adjoint} iff
\begin{equation}
T=T^\dagger, \quad D(T)=D(T^\dagger).
\end{equation}

Let us now illustrate the above concepts using the example of the momentum operator $\hat{p}$ with domain $D(\hat{p})$. We ask ourselves three important questions:
\begin{enumerate}
\item Is the momentum operator $\hat{p}, \ D(\hat{p})$ symmetric?
\item What is its adjoint, more importantly what is $D(\hat{p}^\dagger)$?
\item Is the momentum operator $\hat{p}$ self-adjoint, i.e. is $D(\hat{p})=D(\hat{p}^\dagger)$ fulfilled?
\end{enumerate}
To answer this questions it is very convenient to  define the  following quantity
\begin{equation}\begin{split}\label{pomd}
\delta_{\hat{p}}&:= (\xi, \hat{p}\eta)-(\hat{p}\xi,\eta)=-i\left(\int^{1}_{0}\xi^*\frac{\d \eta}{\d x}\d x-\int^{1}_{0}\frac{\d\xi^*}{\d x}\eta\d x\right)\\
&=-i\left[\xi^*(1)\eta(1)-\xi^*(0)\eta(0)\right],
\end{split}\end{equation}
where $\xi,\eta$ are arbitrary absolutely continuous elements of the Hilbert space $L^2[0,1]$. To check whether $\hat{p}$ is a symmetric operator in the domain $D(\hat{p})$ it is sufficient to evaluate \eqref{pomd} for $\xi,\eta\in D(\hat{p})$. Since \eqref{D(p)} defines the domain, this implies $\xi(1)=\xi(0)=\eta(1)=\eta(0)=0$ and we have
\begin{equation}
\delta_{\hat{p}}=-i\left[\xi^*(1)\eta(1)-\xi^*(0)\eta(0)\right]=0, \quad \forall\xi,\eta\in D(\hat{p}),
\end{equation}
and we conclude that $\hat{p}$ is symmetric in $D(\hat{p})$.\\

The second question is to determine the adjoint, i.e. we have to find what is the domain  of the adjoint operator $D(\hat{p}^\dagger)$. In doing so, let as consider $\eta\in D(\hat{p})$ and ask what are all possible $\xi$ such that $\delta_{\hat{p}}=0$. Since $\eta\in D(\hat{p})$ implies $\eta(1)=\eta(0)=0$, it is easy to see that $\delta_{\hat{p}}=0$ will be  satisfied for any $\eta\in D(\hat{p})$ without any particular condition on $\xi$, apart from being an absolute continuous element of the Hilbert space. This means
\begin{equation}\label{D(pk)}
D(\hat{p}^\dagger)=\left\{\psi | \psi\ \text{absolutely continuous and}\ \psi\in L^2[0,1]\right\}.
\end{equation}

We now address  the third question. Is the momentum operator $\hat{p}$ on the domain $D(\hat{p})$ self-adjoint, i.e. is $D(\hat{p})=D(\hat{p}^\dagger)$?  We immediately see that by comparing \eqref{D(p)} and \eqref{D(pk)}  the answer is negative, and furthermore we see that $D(\hat{p})\subset D(\hat{p}^\dagger)$ as subsets of the Hilbert space $\mathcal{H}$. The conclusion is that the momentum operator  defined by \eqref{p} and \eqref{D(p)} is not a SA operator! An immediate consequence of this statement can be seen in its spectrum. Namely, let us recall that if an operator is SA, then it must have a real spectrum. It is straightforward to verify that the eigenvalue equation for our momentum operator, i.e.
\begin{equation}
\hat{p}\psi=p\psi, \quad \psi(1)=\psi(0)=0,
\end{equation}
has no real solutions for the eigenvalue $p$, illustrating the fact that $\hat{p}, \ D(\hat{p})$ is not a self-adjoint operator.\\

The above discussion shows that the momentum operator $\hat{p}$ with the simple domain \eqref{D(p)} is not a SA operator. This happens due to the fact that the domain of the adjoint operator turns out to be much bigger that the domain of the operator itself. The natural next question to ask is if there exists some other domain in which $\hat{p}$ could be self-adjoint? In order to investigate this question more, let us, for a moment, consider another domain for $\hat{p}$, namely let us define a new domain
 \begin{equation}\label{Dtheta}
D_{\theta}(\hat{p})=\left\{\psi(1)=\text{e}^{\text{i}\theta}\psi(0),\ \psi\ \text{absolutely continuous and}\ \psi\in L^2[0,1]\right\},
\end{equation}
where $\theta\in\mathbb{R}\ \text{mod}\ 2\pi$. We also  assume $\psi(0)\neq 0$, otherwise \eqref{Dtheta}  could reduce to \eqref{D(p)}. We are again  in a position to ask the  three questions about symetricity, adjoint and self-adjointness for the momentum operator in the domain $D_{\theta}(\hat{p})$.\\

In order to check whether  $\hat{p}$ is symmetric in $D_{\theta}(\hat{p})$, we consider $\xi,\eta\in D_{\theta}(\hat{p})$ in \eqref{pomd}. We get
\begin{equation}
\delta_{\hat{p}}=-i\left[\xi^*(1)\eta(1)-\xi^*(0)\eta(0)\right]=-i\left[\text{e}^{-\text{i}\theta}\xi^*(0)\text{e}^{\text{i}\theta}\eta(0)-\xi^*(0)\eta(0)\right]=0, \quad \forall\xi,\eta\in D_{\theta}(\hat{p}),
\end{equation}
concluding that $\hat{p}$ is symmetric in $\xi,\eta\in D_{\theta}(\hat{p})$.\\

Now we want to find the domain of the adjoint. To do so, we have to consider $\eta\in D_{\theta}(\hat{p})$ and find all the  possible $\xi$ so that $\delta_{\hat{p}}=0$.  Therefore in  order to find  the domain of the adjoint we have to find all $\xi\in\mathcal{H}$ by solving the equation
\begin{equation}
\delta_{\hat{p}}=-i\left[\xi^*(1)\eta(1)-\xi^*(0)\eta(0)\right]=0.
\end{equation}
Since $\eta\in D_{\theta}(\hat{p})$ implies $\eta(1)=\text{e}^{\text{i}\theta}\eta(0)$ we get
\begin{equation}
\eta(0)\left[\xi^*(1)\text{e}^{\text{i}\theta}-\xi^*(0)\right]=0\ \Rightarrow\  \xi(1)=\text{e}^{\text{i}\theta}\xi(0)\ \ \ \text{as}\ \ \eta(0)\neq 0.
\end{equation}
Namely, we have for the domain of the adjoint
\begin{equation}
D_{\theta}(\hat{p}^\dagger)=\left\{\psi(1)=\text{e}^{\text{i}\theta}\psi(0),\ \psi\ \text{absolutely continuous and}\ \psi\in L^2[0,1]\right\}.
\end{equation}
It is now obvious that  $D_{\theta}(\hat{p}^\dagger)= D_{\theta}(\hat{p})$ and this answers the third question concluding that $\hat{p}$ is indeed SA in the domain $D_{\theta}(\hat{p})$.\\

We can also look at the  eigenvalue equation for $\hat{p}, D_{\theta}(\hat{p})$  
\begin{equation}
\hat{p}\psi=p\psi, \quad \psi(1)=\text{e}^{\text{i}\theta}\psi(0),
\end{equation}
and find square integrable solutions
\begin{equation}
\psi_{n}(x)=\frac{\text{e}^{\text{i}p x}}{\sqrt{2}}, \quad p=2n\pi+\theta,
\end{equation}
where $n\in\mathbb{Z}$ and $\left\{\psi_n\right\}^{+\infty}_{-\infty}$ form an orthonormal basis in $L^2[0,1]$. Notice that for each value of the parameter $\theta$, the spectrum is different, that is, we have a one parameter family of inequivalent quantizations of the momentum operator on a finite interval.

\subsection{The von Neumann's method: self-adjoint  extensions of a symmetric operator}

In the previous subsection we saw that the momentum operator is not SA on the domain \eqref{D(p)}, but with a suitable choice of domain, that is \eqref{Dtheta}, it can be made SA. However, the domain \eqref{Dtheta} was introduced without any specific physical or mathematical motivation. In this subsection we will introduce the method of SA extensions (a method due to von Neumann) and show that the domain \eqref{Dtheta} is indeed the SA extension of \eqref{D(p)}.\\

Let us start with a given symmetric operator $T$ with a domain $D(T)$ and pose the following questions:
\begin{enumerate}
\item Is the symmetric operator $T$ self-adjoint in $D(T)$?
\item If it is not SA, can it be made self-adjoint?
\item If it can be made SA, what is the suitable domain of self-adjointness?
\end{enumerate}
To answer these questions, we shall not give the derivation or proofs\footnote{For more details see \cite{sae} and Appendix F.} of the results on the SA extension method, but rather use the theorems and illustrate them on examples.\\

Some further definitions are needed before we can better capture the notion of self-adjlointness and its failure.  Given a symetric operator $T$, let us consider  the following equations\footnote{Actually we should consider $T^\dagger \psi_{\pm}=\pm i\kappa \psi_{\pm}$, where $\kappa\in\mathbb{R}$ due to dimensionality reasons (namely $[T]=[\kappa]$), but this we omit here and set $\kappa=1$ as our ``natural'' choice of units. However the dimensionality of $\kappa$ is very important and its physical meaning is that this is the scale of the anomaly, i.e. the breaking of the classical symmetry and it somewhat governs the physically allowed types of interaction on the boundary.}
\begin{equation}\label{pm}
T^\dagger\psi_{\pm}=\pm i\psi_{\pm}, \quad \psi_{\pm}\in D(T^\dagger)
\end{equation}
and let the natural number $n_{\pm}$ denote the number of linearly independent square integrable solution of \eqref{pm}. The pair of natural numbers $(n_{+},n_{-})$ are called deficiency indices for the operator $T$. The deficiency indices are a measure of the deviation of operator $T$ from self-adjointness. Namely, notice that for a truly SA operator, the deficiency indices should be equal to zero, because the spectrum of such an operator consists only of real numbers. This notion is used to classify operators in terms of their deficiency indices. In general, there can be three classes of operators:
\begin{enumerate}
\item $T$ is (essentially) SA iff $(n_{+},n_{-})=(0,0)$, or we say that it has a unique SA extension.
\item If $n_{+}=n_{-}$, then the operator $T$ is not SA but admits SA extensions.
\item In case that $n_{+}\neq n_{-}$, then the operator $T$ is not SA and has no SA extensions.
\end{enumerate}
Let us see what this all means for our momentum operators from the previous subsection. First we look at the operator $\hat{p}$ with the domain $D(\hat{p})$. We showed that $D(\hat{p}^\dagger)$ is given by \eqref{D(pk)}. To determine the deficiency indices, we need to solve the equations
\begin{equation}
-i\frac{\d \psi_{\pm}}{\d x}=\pm i\psi_{\pm}, \quad \psi_{\pm}(x)\in D(\hat{p}^\dagger).
\end{equation}
The square integrable solutions (in the interval $[0,1]$) of the above equations are given by
\begin{equation}\begin{split}\label{Cpm}
&\psi_{+}(x)=C_+\text{e}^{-x},\ \Rightarrow\ n_+=1,\\
&\psi_{-}(x)=C_-\text{e}^{x},\ \Rightarrow\ n_-=1,
\end{split}\end{equation}
where $C_\pm$ are normalization constants. This means that for the deficiency indices we have  $n_+=n_-=1$, which agrees with the  fact that $\hat{p}$ is not SA in $D(\hat{p})$ but now it has SA extensions. On the other hand, what are the deficiency indices for the momentum operator $\hat{p}$ on the domain $D_{\theta}(\hat{p})$ as in \eqref{Dtheta}? Now, we have $D_{\theta}(\hat{p})=D_{\theta}(\hat{p}^\dagger)$ and  the deficiency indices are obtained by solving the equations
\begin{equation}
-i\frac{\d \psi_{\pm}}{\d x}=\pm i\psi_{\pm}, \quad \psi_{\pm}(1)=\text{e}^{i\theta}\psi_{\pm}(0).
\end{equation}
These equations have no square integrable solutions for real values of the parameter $\theta$, implying that $n_+=n_-=0$, i.e. $\hat{p}$ is essentially SA in $D_{\theta}(\hat{p})$. This also agrees with what we concluded earlier.\\

We are now interested in analyzing the second type of operators, namely,  symmetric operator $T$ in a domain $D(T)$ with deficiency indices $n_+=n_-=n$. Such $T$ is not SA but admits SA extensions. Therefore, there exist a suitable domain, or better to say, a suitable  extension of its domain that would render the operator in question SA. This is the main result of the von Neumann's method\footnote{See \cite{sae} and Appendix F.}. Here we state the final result of the method, i.e. the domain of self-adjoitness for a symmetric operator $T$ is given by\footnote{Here the domain is given a bit wage, since one might ask how a finite dimensional matrix $U$ acts on $\psi_{-}$. The details are given in Appendix F, and here our wage definition is good enough since in all of our examples the matrix $U$ will be a pure phase.}
\begin{equation}\label{domain}
D_{U}(T)=\left\{\psi+\psi_{+}+U\psi_{-}|\ \psi\in D(T)\ \text{and}\ \ U \ \text{is an}\ \ n\times n \  \text{unitary matrix}\right\}.
\end{equation}

Now let us go back to our momentum operators and show that \eqref{Dtheta} is indeed the SA extension of \eqref{D(p)}.
For the case of momentum operator, let $\xi(x)$ denote an arbitrary element of the domain in \eqref{domain}. Our task now is to show that $\xi(x)$ constructed from the above  prescription \eqref{domain} obeys the boundary condition $\xi(1)=\text{e}^{i\theta}\xi(0)$, where $\theta\in\mathbb{R}\ \text{mod}\ 2\pi$. This would illustrate the above claim about  the domain of self-adjointness, as we have already  seen  that for  wavefunctions obeying such boundary conditions the momentum operator is indeed SA. In order to proceed, we first rewrite \eqref{Cpm} with normalizations given explicitly:
\begin{equation}\label{exp}
\psi_{+}=\frac{\sqrt{2}\text{e}}{\sqrt{\text{e}^2 -1}}\text{e}^{-x}, \ \ \ \psi_{-}=\frac{\sqrt{2}}{\sqrt{\text{e}^2 -1}}\text{e}^{x}.
\end{equation}
Now note that in our case of the momentum operator, $U$ is a $1\times 1$ matrix, i.e. it is a phase $\text{e}^{i\gamma}$, where $\gamma\in\mathbb{R}$. Now if $\xi(x)$ is an arbitrary element of the domain $D_{\gamma}(\hat{p})$, then from \eqref{domain} we have
\begin{equation}
\xi(x)=\psi(x)+\psi_+(x)+\text{e}^{i\gamma}\psi_-(x),
\end{equation}
where $\psi(x)\in D(\hat{p})$, i.e. $\psi(1)=\psi(0)=0$ and $\psi_{\pm}$ are given by \eqref{exp}. Using this we find that
\begin{equation}\begin{split}\label{skoro}
&\xi(1)=\psi_{+}(1)+\text{e}^{i\gamma}\psi_{-}(1),\\
&\xi(0)=\psi_{+}(0)+\text{e}^{i\gamma}\psi_{-}(0).
\end{split}\end{equation}
Using \eqref{exp} and \eqref{skoro} we get
\begin{equation}
\frac{\xi(1)}{\xi(0)}=\frac{1+\beta\text{e}}{\text{e}+\beta}, \ \ \beta=\text{e}^{i\gamma}, \ \ \ \left|\beta\right|^2=1
\end{equation}
and taking the modulus yields
\begin{equation}
\left|\frac{\xi(1)}{\xi(0)}\right|^2=1\ \Rightarrow\ \xi(1)=\text{e}^{i\theta}\xi(0),
\end{equation}
where $\theta\in\mathbb{R}\ \text{mod}\ 2\pi$. This illustrates the claim \eqref{domain}.

\subsection{Motion of a free particle on $\left[0,+\infty\right)$}
It is easy to see that the momentum operator defined on a whole half line $\mathbb{R}_+=\left[0,+\infty\right)$ with the boundary condition $\psi(0)=0$ and $\psi\in L^2(\mathbb{R}_+)$ has unequal deficiency indices.  Namely, solve \eqref{pm} and we get for $\psi_\pm$ exactly \eqref{Cpm} again, but since we are looking for square integrable solution on $\mathbb{R}_+$ we get $n_+ =1$ and $n_{-}=0$. By the von Neumann theorem, this implies that the momentum operator on the half line can never be realized as a SA operator in QM\footnote{For a new concept for the quantization of momentum operator in space with boundaries see \cite{newc}.}! \\

Even though the momentum operator fails to be an observable of the free particle on a half line, this is not the case for the Hamiltonian. Namely, let us consider the Hamiltonian for a free particle\footnote{For simplicity we set the  mass to be $m=\frac{1}{2}$.} on a half-line  $x\in\mathbb{R}_+$ as
\begin{equation}\label{H}
H=-\frac{\d^2}{\d x^2}.
\end{equation}
We also need a domain to properly define an operator in Hilbert space and for that matter we consider a subset $d(H)$ of the Hilbert space $\mathcal{H}$ given by
\begin{equation}
d(H)=\left\{\psi\ |\ \psi\in L^2(\mathbb{R}_{+}), H\psi\in L^2(\mathbb{R}_{+}), \psi^\prime\ \text{absolutely continuous}\right\}
\end{equation}
and using it we define the domain of the Hamiltonian  $H$ as 
\begin{equation}\label{D(H)}
D(H)=\left\{\psi\ |\ \psi\in d(H),\ \psi(0)=\psi^\prime(0)=0\right\}.
\end{equation}
From the physical stand-point, the domain \eqref{D(H)} looks natural since it reflects the fact that at zero we have an impenetrable wall.

But now, as we did for the momentum operator before, we have to answer the following three questions:
\begin{enumerate}
\item Is the Hamilton operator $H$ symmetric in $D(H)$?
\item What is the domain of its adjoint $D(H^\dagger)$?
\item Is the operator $H$ SA in the domain $D(H)$?
\end{enumerate}
Again, it is convenient to examine the following quantity
\begin{equation}\begin{split}
\delta_{H}&:= (\xi,H\eta)-(H\xi,\eta)=-\int^{\infty}_{0}\left(\xi^*\frac{\d^2 \eta}{\d x^2}-\frac{\d^2 \xi}{\d x^2}\eta\right)\d x\\
&=\xi^*(0)\frac{\d\eta(0)}{\d x}-\frac{\d\xi^*(0)}{\d x}\eta(0),
\end{split}\end{equation}
where $\xi,\eta$ are, at the moment, general elements of the Hilbert space $\mathcal{H}$ which vanish at infinity, i.e. they are square integrable. To see whether  $H$ is symmetric or not in $D(H)$, we have to consider $\xi,\eta\in D(H)$. In doing so, using \eqref{D(H)}, i.e. $\xi(0)=\xi^{\prime}(0)=\eta(0)=\eta^\prime(0)=0$, we obtain
\begin{equation}
\delta_{H}=\xi^*(0)\frac{\d\eta(0)}{\d x}-\frac{\d\xi^*(0)}{\d x}\eta(0)=0, \quad \forall\xi,\eta\in D(H),
\end{equation}
and we conclude that the operator $H$ is symmetric in $D(H)$.\\

Now we proceed to determine the domain of the adjoint $D(H^\dagger)$. In order to do so, we have to consider $\eta\in D(H)$ and find  all possible $\xi$ such that $\delta_{H}=0$. Since $\eta\in D(H), \eta(0)=\eta^\prime (0)=0$, we see that $\delta_{H}=0$ is satisfied for any $\eta\in D(H)$ without any particular condition on $\xi$. We conclude that the domain of the adjoint is defined by
\begin{equation}
D(H^\dagger)=\left\{\xi\ | \ \xi\in d(H)\right\}.
\end{equation}
\\

Note that, as in the case of the momentum operator on a finite interval, here also $D(H)\subset D(H^\dagger)$ as subsets of $\mathcal{H}$, implying that the operator $H$ is not SA in $D(H)$.\\

It is now indicative that the failure of self-adjointness, leading to  $D(H^\dagger)\supset D(H)$ is a consequence of defining the domain $D(H)$ too restrictively. We may expect that changing the boundary conditions, in order to enlarge $D(H)$ may render the operator $H$ self-adjoint. With this in mind, let us examine the domain where we have a Robin-type boundary condition
\begin{equation}\label{Dalpha}
D_{\alpha}(H)=\left\{\psi\ |\ \psi\in d(H), \psi^\prime(0)=\alpha\psi(0)\right\},
\end{equation}
where $\alpha\in\mathbb{R}$. It is straightforward to see that for this domain we have the following properties:
\begin{enumerate}
\item  Operator $H$ is symmetric in $D_{\alpha}(H)$.
\item $D_{\alpha}(H^\dagger)=D_{\alpha}(H)$.
\item Concluding that $H$ is SA in $D_{\alpha}(H)$.
\end{enumerate}
This shows that by enlarging the domain of $H$ has reduced the domain of $H^\dagger$ in such a way that they are now equal and the operator is SA.\\

Let us now go back to our Hamiltonian defined on a domain $D(H)$ and  proceed with the method of von Neumann. First, we need to find the deficiency indices. In order to do so, we have to find square integrable solutions of the equations
\begin{equation}\label{defind}
-\frac{\d^2 \psi_{\pm}}{\d x^2}=\pm i\psi_{\pm}, \quad \forall \psi_{\pm}\in D(H^\dagger).
\end{equation}
The  normalized solutions are given by
\begin{equation}\begin{split}\label{resH}
&\psi_{+}=2^{1/4}\text{exp}\left(\frac{(i-1)}{\sqrt{2}}x\right) \ \Longrightarrow\ n_+=1\\
&\psi_{-}=2^{1/4}\text{exp}\left(\frac{-(i+1)}{\sqrt{2}}x\right) \ \Longrightarrow\ n_-=1
\end{split}\end{equation}

Using the prescription \eqref{domain} of von Neumann, we can find the domain of self-adjointness of $H$. From \eqref{domain} we expect that the SA domain of $H$ is 
\begin{equation}\label{Dbeta}
D_{\beta}(H)=\left\{\xi\ |\ \xi=\psi+\psi_++\beta\psi_-\right\},
\end{equation}
where $\psi\in D(H)$, $\psi_\pm$ are given in \eqref{resH} and $\beta=\text{e}^{i\gamma}$ is a unitary $1\times 1$ matrix, i.e. a pure phase. It can be shown that if $\xi\in D_{\beta}(H)$, then
\begin{equation}\label{bound}
\left|\frac{\xi^\prime(0)}{\xi(0)}\right|^2=\frac{1-\sin\gamma}{1+\cos\gamma}\ \Longrightarrow\ \xi^\prime(0)=\alpha\xi(0)
\end{equation}
where\footnote{And the parametrization $\alpha=\pm\frac{\cos(\gamma/2+\pi/4)}{\cos(\gamma/2)}$ is used.} $\alpha\in\mathbb{R}$, which is the same as the condition on the wavefunction as given by \eqref{Dalpha}, i.e. the domains given by \eqref{Dalpha} and \eqref{Dbeta} are identical. Thus we again have an illustration of the prescription \eqref{domain} of von Neumann. Again, we can conclude that a quantum free particle on a half line is described by a one parameter family of inequivalent Hamiltonians. It is up to physics, that is experiment, to ``choose'' the description, i.e. for which parameter $\alpha$ we have the best agreement with experimental data.\\

From now on, we shall no longer illustrate \eqref{domain} but shall assume its validity and shall obtain the domain of self-adjointness using it as a rule.
For more illustrative examples see \cite{example}.

\section{Some physical applications}

\subsection{New bound states}
In this subsection let us calculate and investigate the properties of  solutions of the stationary Schr\"{o}dinger equation $H\psi=E\psi$ for the case  of a free particle moving on the semi-interval $\mathbb{R}_+$. The Hamiltonian for this system is given in \eqref{H} and the Schr\"{o}dinger equation  is
\begin{equation}\label{sch}
-\frac{\d^2\psi}{\d x^2}=E\psi, \quad \psi^\prime(0)=\alpha\psi(0).
\end{equation}
Now, the solutions of the equation \eqref{sch} highly depend on the sign\footnote{To see this more closely, act on \eqref{sch} with $(\psi,\cdot)$ to obtain $$E=\frac{\alpha\left|\psi(0)\right|^2+\left\|\psi^\prime\right\|^2}{\left\|\psi\right\|^2}$$ rendering that the energy is real if $\alpha\in\mathbb{R}$ and can be positive or negative depending on $\psi$.} of the energy eigenvalue $E$. Let us first consider the case of positive energy, i.e. $E:= k^2>0$. The most general solution in this case  is given by\footnote{There are no square-integrable solutions of this equation for positive energy. Here one actually looks for a solution of a generalized eigen-equation within the formalism of rigged Hilbert space (see Appendix G for further comments.)}
\begin{equation}
\psi=C\text{e}^{-ikx}+D\text{e}^{ikx},
\end{equation}
where  $C,D$ are constants that are jet to be determined.  After imposing the  boundary condition \eqref{sch} we obtain
\begin{equation}
\frac{D}{C}=\frac{ik+\alpha}{ik-\alpha}
\end{equation}
and the solution can be written in the following form
\begin{equation}\label{scat}
\psi=C\left(\text{e}^{-ikx}+\frac{ik+\alpha}{ik-\alpha}\text{e}^{ikx}\right).
\end{equation}
Since \eqref{scat} are not elements of the Hilbert space $L^2(\mathbb{R}_{+})$ we do not normalize them to 1, but rather interpret them as ``stream'' of particles. Therefore we have  streams of particles with momentum $k$ and $-k$ both with their corresponding amplitude. Therefore, we can define the so called reflection coefficient as $R:= \frac{ik+\alpha}{ik-\alpha}$, and it is easy to see that it satisfies $RR^*=1$, i.e.  the reflection coefficient is a pure phase and we can write it as $R=\text{e}^{-i\theta(\alpha)}$. Here $\theta(\alpha)$ is the phase of the reflected wave that depends on the parameter $\alpha$ that controls the allowable types of interaction on the boundary. The solutions \eqref{scat} are often called  scattering states. This is often the case in physics. Namely, when solving the Schrodinger equation for some localized potential it is often the case that one has a set of solutions for positive energy (often this is the continuous part of the spectrum) and another set of solutions for negative energies (often the point spectrum). The positive energy solutions are generalizations of plane waves, they are not elements of the Hilbert space in question (not square integrable), but rather distributions and represent the scattering states of the system, while the negative energy solutions are elements of the Hilbert space  and are often called bound states.  \\

We now consider the negative energy solutions, i.e. we take $E:= -\mathcal{E}$, $\mathcal{E}>0$ and look for square integrable solutions of Schr\"{o}dinger's equation \eqref{sch}. In doing so, we use an ansatz
\begin{equation}\label{ansatz}
\psi=D\text{e}^{\lambda x}, \quad \lambda, D\in\mathbb{R}.
\end{equation}
 The equation \eqref{ansatz} together with \eqref{sch} and $E:= -\mathcal{E}$ lead to,
\begin{equation}\label{ansatz2}
\mathcal{E}=\lambda^2 \ \Longrightarrow \lambda_{1,2}=\pm\sqrt{\mathcal{E}} \ \Longrightarrow\ \psi=D_1\text{e}^{\sqrt{\mathcal{E}} x}+D_2\text{e}^{-\sqrt{\mathcal{E}} x},
\end{equation}
where $D_1=0$ due to square-integrability. 
Imposing the boundary condition \eqref{sch} on the wavefunction \eqref{ansatz2} we obtain
\begin{equation}
\alpha=-\sqrt{\mathcal{E}}.
\end{equation}
Since $\mathcal{E}>0$ implies $\alpha<0$, this insures the square integrability of our solution, i.e. bound state. Finally, the complete solution of the bound state eigenvalue problem can be written in the following way
\begin{equation}\label{bound1}
\psi(x)=\sqrt{2\left|\alpha\right|}\text{e}^{\alpha x}, \quad \alpha<0, \quad E=-\alpha^2.
\end{equation}
At this stage we pause and ask the question how is it possible for a free particle to admit a bound state. Namely, for a bound state to exist, one needs some scale, which is not apparent in the free Hamiltonian. Since there is no scale in the Hamiltonian, it should be scale invariant and all the solutions could be scaled to zero momentum $k$, namely illustrating that the Hamiltonian in question just has the continuous part of the spectrum. However, the scale is hidden in the domain of the SA Hamiltonian, i.e. the boundary condition \eqref{sch} which defines the domain of self-adjointness supplies the necessary scale. Notice that the parameter $\alpha$ in \eqref{sch} has dimensions of inverse length and is the relevant dimensionful parameter related to the bound state energy \eqref{bound1}. One might be puzzled by the question on how does a dimensionful scale appear in the general method of von Neumann. In order to answer this, note that  equations \eqref{defind} for $\psi_{\pm}$
must have a constant with dimension length squared on the right side, whose value has conveniently been set equal to unity. Therefore, the dimensionful constant  actually enters the  very definition of the domain, i.e. the boundary condition through \eqref{defind}. The existence of such a  dimensionful constant breaks the scale invariance of the system in question and in our case produces the bound state when $\alpha<0$. This is perhaps the simplest example of quantum mechanical breakdown of scale invariance, i.e. the manifestation of the so called scale anomaly. Examples that have this exact feature are  the systems with the $\delta$-function potential and inverse-square potential \cite{Cabo}.

\subsection{Anomalies in Hamiltonian formalism}

Symmetries of a classical system could be broken in the process of quantization. This is due to the so called anomalies \cite{tasi}. This was first realized when studying quantum field theories in a perturbative approach \cite{pert}. Since anomalies are usually characterized by the geometric and topological features of the theory \cite{geom}, it was soon realized that they are fundamentally non-perturbative in nature \cite{nonp}. It was exatly this point of view that ultimately led to a variety of applications of the theory of anomalies in areas ranging from black hole \cite{abh, jos} and particle physics \cite{part}, all the way to condensed matter physics and quantum Hall effect \cite{hall}.\\

In order to understand anomalies one needs to realize that observables (like Hamiltonian, momentum etc.) are often unbounded operators \cite{sae}. To have a well defined unbounded operator as an observable in QM one needs to define its domain of self-adjointness using appropriate boundary conditions. The symmetry of a system is implemented by the fact that the corresponding generator of symmetry commutes with the Hamiltonian and preserve its domain. In case that the symmetry generator does not preserve the domain of the Hamiltonian, we say that the symmetry is broken due to quantization and anomaly appears. To see this better, let $H$ be the Hamiltonian of some system and $A$ be a generator of symmetry. Usually in standard textbooks on QM for physicist it is stated that $A$ is a symmetry iff $[H,A]=0$ without any reference to the domain of operators. But the operator is defined by its rule of acting and domain. So, we have to consider operators as pairs $(H, D(H))$ and $(A, D(A))$ when evaluating the commutator, that is we have to act with the commutator on some element of Hilbert space
\begin{equation}\label{commute}
[H,A]\psi=(HA-AH)\psi=H(A\psi)-A(H\psi).
\end{equation}
For $\psi\in D(H)$  eq. \eqref{commute} makes sense only if $A\psi\in D(H)$ and $H\psi\in D(A)$. If this is not true, then anomaly appears.  That is one has to check how the generator of symmetry $A$ acts on the domain of the Hamiltonian $H$ and see if it leaves it invariant. If yes, we say that $A$ is a true symmetry, if not then $A$-symmetry is broken due to quantization and $A$ is anomalous.\\

To illustrate this, let us first derive the Heisenberg equation of motion how it is usually presented in standard QM textbooks, that is by omitting the subtle domain issues. Let $A$ be an observable represented by a SA operator, $\psi\equiv\psi(t)\in L^2(\mathbb{R})$ satisfying the Schr\"{o}dinger equation
\begin{equation}\label{a1}
i\frac{\partial}{\partial t}\psi=H\psi.
\end{equation}
The expectation value of $A$ in a state $\psi$ is defined by
\begin{equation}\label{a2}
\left\langle A\right\rangle_{\psi}=(\psi,A\psi).
\end{equation}
Now we take the time derivative $\frac{\d}{\d t}$ of \eqref{a2} and using \eqref{a1} we have
\begin{equation}\begin{split}\label{a3}
\frac{\d \left\langle A\right\rangle_{\psi}}{\d t} &=(\frac{\partial \psi}{\partial t},A\psi)+(\psi,\frac{\partial A}{\partial t}\psi)+(\psi,A\frac{\partial \psi}{\partial t})  \\
&=(-iH\psi,A\psi)+\left\langle \frac{\partial A}{\partial t}\right\rangle_{\psi}+(\psi,-iAH\psi)  \\
&=\left\langle \frac{\partial A}{\partial t}\right\rangle_{\psi}+i(\psi,HA\psi)-i(\psi,AH\psi)\\
&=\left\langle \frac{\partial A}{\partial t}\right\rangle_{\psi}+i\left\langle [H,A]\right\rangle_{\psi},
\end{split}\end{equation}
which represents the Heisenberg equation of motion for the expectation values of $A$ and often it is assumed that \eqref{a3} is valid $\forall\psi\in L^2(\mathbb{R})$ so it can be written as an operator equation
\begin{equation}\label{a4}
\frac{\d  A}{\d t}=\frac{\partial A}{\partial t}+i[H,A].
\end{equation}
If $\frac{\d  A}{\d t}=0$ it means that $\left\langle A\right\rangle_{\psi}$ are constant $\forall t$ and $A$ is called a generator of a symmetry. Often we have situations in which $A$ does not explicitly depend on time, $\frac{\partial A}{\partial t}=0$, so the claim $[H,A]=0$ is equivalent to saying $A$ generates a symmetry.\\

However, equation \eqref{a4} and conclusions deduced from it are only valid under very strong assumptions, that were overseen in the above derivation. Namely, already in \eqref{a1} one assumes $\psi\in D(H)$, in \eqref{a2} $\psi\in D(A)$ is assumed, and when further calculating \eqref{a3} and forming the commutator $[H,A]$ one assumes $H\psi\in D(A)$ and $A\psi\in D(H)$. Also $\psi\in D(\frac{\partial A}{\partial t})$ is assumed. Actually, the commutator $[H, A]$ and \eqref{a4} is well defined only if $\psi\in\mathcal{T}$, where $\mathcal{T}$ is a subspace of $L^2(\mathbb{R})$ which is invariant under the action of $H$ and $A$. Namely, $\mathcal{T}$ has to be the intersection of all the powers of $H$ and $A$
\begin{equation}
\mathcal{T}:=\bigcap^{\infty}_{n,m=0}D(H^n A^m).
\end{equation}
$\mathcal{T}$ is called the maximal invariant subspace of the algebra generated by $H$ and $A$ because $A\mathcal{T},H\mathcal{T}\subset\mathcal{T}$ and on this space all the expectation values, uncertainties etc. are well defined. Unfortunately, physicists really just need $\psi\in D(H)$ (or $\psi\in D(H)\cap D(A)$) and are interested only in solutions of \eqref{a1} and therefore encounter possible problems, that is anomalies when using \eqref{a3} and \eqref{a4}.\\

In order to quantify the correction to the Heisenberg equation of motion \eqref{a4}, let us look carefully at the second line in \eqref{a3}, $\frac{\d \left\langle A\right\rangle_{\psi}}{\d t} =\left\langle \frac{\partial A}{\partial t}\right\rangle_{\psi}+i(H\psi,A\psi)-i(\psi,AH\psi)$ and concentrate on the last two terms. The term $(H\psi,A\psi)$ is well defined as long as $\psi\in D(H)\cap D(A)$ while the last term $(\psi,AH\psi)$ has a problem if $H\psi\notin D(A)$. If the algebraically\footnote{This is usually the case, for example the canonical commutation relation $[\hat{x},\hat{p}]=i\hbar I$, where the right hand side is algebraically obtained and, as an operator, is well defined on all of $L^2(\mathbb{R})$. Algebraically obtained would here mean that we calculated $[H,A]$ evaluated on $\mathcal{T}$. } obtained commutator $[A,H]$ is well defined on all $L^2(\mathbb{R})$ then we could  write $AH=[A,H]+HA$ to obtain
\begin{equation}
(H\psi,A\psi)-(\psi,AH\psi)=(\psi,[H,A]\psi)-i\mathfrak{A}
\end{equation}
where
\begin{equation}\label{ano}
\mathfrak{A}:=i\left((H\psi,A\psi)-(\psi,HA\psi)\right)=i\left\langle \right(H^{\dagger}-H)A\rangle_{\psi}.
\end{equation}
Taking the above in consideration, we can rewrite the Heisenberg equation of motion as
\begin{equation}\begin{split}\label{H1}
\frac{\d \left\langle A\right\rangle_{\psi}}{\d t} &=\left\langle \frac{\partial A}{\partial t}\right\rangle_{\psi}+i\left\langle [H,A]\right\rangle_{\psi}+\mathfrak{A},\\
\frac{\d  A}{\d t}&=\frac{\partial A}{\partial t}+i[H,A]+i(H^\dagger-H)A.
\end{split}\end{equation}
Of course $\mathfrak{A}$ is called the anomaly \cite{anomalies}. We see that whenever $A$ keeps the domain $D(H)$ invariant, $AD(H)\subset D(H)$ and $H^\dagger =H$ on $D(H)$, then the anomaly vanishes $\mathfrak{A}=0$. But whenever $A$ does not keep the domain $D(H)$ invariant, $A\psi\notin D(H)$, the extra term in \eqref{H1} will produce a non-zero surface contribution responsible for the anomaly. We can conclude the following. In the presence of the anomaly, the commutator $[H,A]$ in \eqref{a4} has two contributions, the regular and anomalous part, $[H,A]=[H,A]_{reg}+[H,A]_{\mathfrak{A}}$. The regular part is the extension of the algebraically obtained commutator to the whole Hilbert space, while the anomalous part is $(H^\dagger-H)A$. It can be shown that the anomaly introduced here is equivalent to the one in the path integral approach \cite{anomalies, Fuji1, Fuji2}. This general approach to anomalies \cite{delta, anomalies, an1} can be applied in various systems like integrable models \cite{integrable}, condensed matter physics \cite{qhe}, molecular and atomic physics \cite{molec}, and even black hole physics \cite{sen}. In \cite{an1, aqft} it is argued that this approach can be adopted in quantum field theories as well, since the information about the domain of self-adjointness is encoded in the mode expansion of the fields through appropriate boundary conditions.\\

Here we will illustrate the occurrence of the anomaly in our simple system of non-relativistic particle on a half-line (see Section III.C and IV.A). Namely, in  classical physics the Hamiltonian for our system is equal to $H=p^2$ and our system is scale invariant\footnote{Actually it is $SO(2,1)$ invariant and the symmetry is generated by the Hamiltonian $H$, dilatation $D$ and conformal generator $K$.}. To put it more formally, the system is invariant under rescaling $x\longrightarrow a x$ , $a\in\mathbb{R}$ and the generator of the scale symmetry is given by the dilatation\footnote{For more details on scale symmetry in classical physics see Appendix H.} $D=tH-\frac{1}{2}xp$. It is easy to show that
\begin{equation}
\left\{H,D\right\}=H
\end{equation}
which together with the Poisson bracket formalism
\begin{equation}
\frac{\d  D}{\d t}=\frac{\partial D}{\partial t}+\left\{D,H\right\}
\end{equation}
gives
\begin{equation}
\frac{\d  D}{\d t}=0,
\end{equation}
showing that $D$ is conserved and proving that our system is scale invariant.\\

Now we want to look the same system on the quantum level. Using our naive quantization procedure we would promote the classical observables $H$ and $D$ to operators
\begin{equation}\begin{split}
D&\longrightarrow\hat{D}=t\hat{H}-\frac{1}{4}(\hat{x}\hat{p}+\hat{p}\hat{x})\\
H&\longrightarrow\hat{H}=\hat{p}^2=-\frac{\d^2}{\d x^2}
\end{split}\end{equation}
and say that they act on the whole Hilbert space $L^2(\mathbb{R}_+)$ and calculate the Heisenberg equation of motion \eqref{a4} and obtain
\begin{equation}
\frac{\d  \hat{D}}{\d t}=0,
\end{equation}
where we used $i[\hat{D},\hat{H}]=\hat{H}$. This way, the system would again be scale invariant and this leads to the conclusion that there cannot exist a normalizable bound state, which contradicts the result found in previous subsection, namely \eqref{bound1}. However, when deducing the non-existence of bound states we used the assumption \eqref{a4} which is not valid due to domain issues of $\hat{D}$ and $\hat{H}$ and the fact that we are missing the anomaly term \eqref{ano}.\\

In section III.C we have properly defined the Hamiltonian $\hat{H}$ as a SA operator on a domain with suitable boundary conditions \eqref{bound}, namely  $\hat{H}:=(H=-\frac{\d^2}{\d x^2},D_{\alpha}(H))$ and for that operator we were able to find the bound state \eqref{bound1} since the new scale, the scale of symmetry breaking or anomaly came through the necessity of the parameter $\alpha$ which classified allowable SA boundary conditions (that is types of interaction on the boundary). Let us now calculate the anomaly \eqref{ano} for our system. We have
\begin{equation}\label{arac}
\mathfrak{A}=i\left((\hat{H}\psi,\hat{D}\psi)-(\psi,H\hat{D}\psi)\right), \quad \psi(x)=\sqrt{2\left|\alpha\right|}\text{e}^{\alpha x}, \quad \alpha<0.
\end{equation}
The first term in \eqref{arac} is given by
\begin{equation}
(\hat{H}\psi,\hat{D}\psi)=\alpha^4 t +\alpha^2(\psi,\hat{G}\psi)
\end{equation}
where $\hat{H}\psi=-\alpha^2\psi$ and $\hat{G}=\frac{1}{4}(\hat{x}\hat{p}+\hat{p}\hat{x})$. Further
\begin{equation}
(\psi,\hat{G}\psi)=\frac{i\alpha}{2}\int^{\infty}_{0}dx\text{e}^{\alpha x}\left(x\frac{\d }{\d x}+\frac{\d }{\d x}x\right)\text{e}^{\alpha x}=\frac{i\alpha}{2}\int^{\infty}_{0}dx\text{e}^{2\alpha x}(2\alpha x+1)
\end{equation}
leading to 
\begin{equation}\label{r1}
(\hat{H}\psi,\hat{D}\psi)=\alpha^4 t +\frac{i\alpha^3}{2}\int^{\infty}_{0}dx\text{e}^{2\alpha x}(2\alpha x+1),
\end{equation}
while for the second term in \eqref{arac} we have
\begin{equation}
(\psi,H\hat{D}\psi)=\alpha^4 t -(\psi,H\hat{G}\psi),
\end{equation}
where
\begin{equation}
(\psi,H\hat{G}\psi)=-\frac{i\alpha}{2}\int^{\infty}_{0}dx\text{e}^{\alpha x}\frac{\d^2 }{\d x^2}\left(x\frac{\d }{\d x}+\frac{\d }{\d x}x\right)\text{e}^{\alpha x}=-\frac{i\alpha^3}{2}\int^{\infty}_{0}dx\text{e}^{2\alpha x}(2\alpha x+5)
\end{equation}
leading to
\begin{equation}\label{r2}
(\psi,H\hat{D}\psi)=\alpha^4 t+\frac{i\alpha^3}{2}\int^{\infty}_{0}dx\text{e}^{2\alpha x}(2\alpha x+5).
\end{equation}
Combining \eqref{r1}, \eqref{r2} and \eqref{arac} we finally get
\begin{equation}
\mathfrak{A}=2\alpha^3\int^{\infty}_{0}dx\text{e}^{2\alpha x}=-\alpha^2\equiv E,
\end{equation}
so the anomaly is exactly equal to the energy of the bound state \eqref{bound1}. Similar analysis for both the bound states and the scale anomaly can be done for the $\delta$-function potential and $1/x^2$ potential \cite{delta, anomalies, Cabo}. We see that the anomaly in the equation of motion for the dilatation is different from zero since the self-adjointness domain \eqref{Dalpha} of the Hamiltonian is not dilatation invariant. Namely, if $\psi\in D_{\alpha}(H)$ one can check that $D\psi\in D_{\alpha^\prime}(H)$, $\alpha^\prime\neq\alpha$, namely $\psi$ and $D\psi$ are in different SA exstension of the Hamiltonian $H$.

\subsection{Symmetry and degeneracy of the spectrum: circle vs finite interval}

In this subsection we will analyze and confront the properties of apparently two very similar systems: free particle on a circle vs free particle on a finite interval. We will look in detail the corresponding momentum and Hamiltonian operators, its domains of self-adjointness, spectrums and symmetry properties. We will show that, due to domain subtleties, there is an important difference of these two systems and it is related to the fact that the system on a circle  exhibits a translation symmetry, leading to a degenerate spectrum, while the system on finite interval has no such property.\\

First let us take a closer look at the system that consists of a free particle moving on a circle, i.e. $x\in\left[0,L\right)$.  Therefore the appropriate Hilbert space is $L^2[0,L]$ and we are interested in elements, that is wavefunctions, that are periodic 
\begin{equation}\label{period}
\psi(x+L)=\psi(x), \quad \psi\in L^2[0,L],
\end{equation}
since this characterizes the fact that we have a particle on a circle.\footnote{Which is equivalent to a situation where we consider the whole real line, but divide it in segments of length $L$ and demand periodicity, i.e. identify all the points with respect to $\mathbb{R}$ mod $L$.}  We are interested in possible momentum operators
\begin{equation}
\hat{p}=-i\frac{\d}{\d x}
\end{equation}
and by applying the ``maximal ignorance'' approach we investigate allowable domains. First of all, we want that $\hat{p}$ is symmetric, namely
\begin{equation}\label{condp}
(\psi,\hat{p}\varphi)=(\hat{p}\psi,\varphi), \quad \forall \psi,\varphi\in d(\hat{p}).
\end{equation}
This way we define the domain of symmetricity $d(\hat{p})$. Explicitly, the condition \eqref{condp} leads to 
\begin{equation}\label{boundsy}
\psi^*(L)\varphi(L)=\psi^*(0)\varphi(0),
\end{equation}
where we used the inner product and partial integration. In doing so, we tacitly assumed that $\psi, \varphi, \psi^\prime, \varphi^\prime \in L^2[0,L]$ and that they are absolutely continuous. Therefore, the domain that renders the momentum operator $\hat{p}$ symmetric is given by
\begin{equation}\label{sym}
d(\hat{p})=\left\{\psi \ | \ \psi,\psi^\prime\in L^2[0,L] \  \text{and}\ \left|\psi(L)\right|=\left|\psi(0)\right|\right\}.
\end{equation}
The domain \eqref{sym} contains a large class of symetric and even SA operators, due to the fact that the boundary condition \eqref{boundsy} can be realized in various ways (for example: Dirichlet, periodic or twisted periodic conditions). We see that the periodic functions form a subset of the domain \eqref{sym}, i.e. the periodicity condition \eqref{period} is included in \eqref{boundsy}. Therefore, we can define a specific symmetric momentum operator by restricting to the domain of periodic functions, i.e. 
\begin{equation}\label{mSA}
\hat{p}=-i\frac{\d}{\d x}, \quad D(\hat{p})=\left\{\psi \ | \ \psi\in d(\hat{p}),\ \text{and}\  \psi(x+L)=\psi(x)\right\}.
\end{equation}
The domain of its adjoint is determined by all $\psi$ such that
\begin{equation}
\delta_{\hat{p}}=(\psi,\hat{p}\varphi)-(\hat{p}\psi,\varphi)=0, \quad \forall\varphi\in D(\hat{p}), \ \psi\in L^2[0,L].
\end{equation}
This leads to the condition
\begin{equation}
\psi^*(L)=\psi^*(0),
\end{equation}
which for $x\in\left[0,L\right)$ is equivalent to the periodicity condition \eqref{period} and we conclude that
\begin{equation}
D(\hat{p}^\dagger)=D(\hat{p}),
\end{equation}
implying that the momentum operator \eqref{mSA} is indeed SA. In order to further illustrate this let us solve the eigenvalue equation for \eqref{mSA}, i.e.
\begin{equation}\label{eve}
\hat{p}\psi=p\psi, \quad \psi\in D(\hat{p}).
\end{equation}
Before we start solving the differential equation in question, it could be useful to know the nature of the eigenvalue $p$, namely to know if $p\in \mathbb{R}, \mathbb{R}_+$ or $\mathbb{C}$. The symmetricity ensures that $p$ is real-valed. Namely, act by $(\psi,\cdot)$ from the left and with $(\cdot, \psi)$ from the right in order to obtain
\begin{equation}
(\psi,\hat{p}\psi)=p(\psi,\psi) \quad \text{and}\  (\hat{p}\psi,\psi)=p^*(\psi,\psi) \ \overset{(\psi,\hat{p}\psi)=(\hat{p}\psi,\psi,)}{\Longrightarrow} p^*=p,\ \forall \psi\in D(\hat{p}).
\end{equation}
It is straightforward to show that the eigenvalue equation has the following normalized solutions
\begin{equation}
\psi_n(x)=\frac{\text{e}^{ip_n x}}{\sqrt{L}}, \quad p_n=\frac{2n\pi}{L}, \quad n\in\mathbb{Z}.
\end{equation}
The set $\left\{\psi_n\right\}^{+\infty}_{-\infty}$ form an orthonormal basis in $L^2[0,L]$. It is easy to see that the deficiency indices are $n_+=n_-=0$ since the eigenvalue equation \eqref{eve} has no solutions for $p\in i\mathbb{R}$ and $\psi\in D(\hat{p}^\dagger)$. \\

Now let us turn our attention to the Hamiltonian of the system. It is an operator given by\footnote{Where we again take $\hbar=1$ and $m=1/2$ for simplicity.}
\begin{equation}
\hat{H}=-\frac{\d^2}{\d x^2}
\end{equation}
and we are jet to determine an appropriate domain. First let us find the domain of symmetricity. As before, we demand 
\begin{equation}
(\psi,\hat{H}\varphi)=(\hat{H}\psi,\varphi), \quad \forall \psi,\varphi\in d(\hat{H})
\end{equation}
leading to the condition
\begin{equation}
\left(\frac{\d\psi^*}{\d x}\varphi-\psi^*\frac{\d\varphi}{\d x}\right)^{L}_{0}=0.
\end{equation}
Here we had to assume that $\psi, \varphi, \psi^\prime, \psi^{\prime\prime}, \varphi^\prime,$ and  $\varphi^{\prime\prime} \in L^2[0,L]$. Therefore, the domain of symmetricity is given by
\begin{equation}
d(\hat{H})=\left\{\psi \ \left| \ \psi,\psi^\prime, \psi^{\prime\prime}\in L^2[0,L] \ \text{and}  \left.\frac{\d\psi^*}{\d x}\psi \right|^{L}_{0}=\left.\psi^*\frac{\d\psi}{\d x}\right|^{L}_{0}\right.\right\}.
\end{equation}
The boundary conditions here are very general and we see that periodic functions form a subset of this domain. Therefore we will restrict ourselves to a specific symmetric Hamiltonian defined with periodic boundary conditions \eqref{period}. In this case, the domain actually coincides with \eqref{mSA}, i.e.
\begin{equation}
D(\hat{H})=\left\{\psi \ \left| \ \psi\in d(\hat{H}),\  \psi(x)=\psi(x+L)\right.\right\}=\left\{\psi \ \left| \ \psi, \psi^\prime\in D(\hat{p}),\  \psi(x)=\psi(x+L)\right.\right\}=D(\hat{p}).
\end{equation}
This is due to the fact that a derivative of a periodic function is again periodic, provided it exists. It remains to see whether this domain is also the domain of self-adjointness. To answer this we look at the domain of the adjoint by finding all $\psi\in L^2[0,L]$ such that 
\begin{equation}
\delta_{\hat{H}}=(\psi,\hat{H}\varphi)-(\hat{H}\psi,\varphi)=0, \quad \forall\varphi\in D(\hat{H})=D(\hat{p}), \ \psi\in L^2[0,L]
\end{equation}
leading to 
\begin{equation}
\psi(L)=\psi(0), \quad \text{and} \quad \frac{\d \psi}{\d x}(L)=\frac{\d \psi}{\d x}(0),
\end{equation}
meaning that $\psi\in D(\hat{H}^\dagger)$ are also periodic functions, implying
\begin{equation}
D(\hat{H})=D(\hat{H}^\dagger)
\end{equation}
and concluding that the Hamiltonian in question is indeed a SA operator. Now let us proceed with analyzing the eigenvalue equation, i.e. the Schr\"{o}dinger equation
\begin{equation}\label{schh}
\hat{H}\psi=E\psi, \quad \psi\in D(\hat{H}).
\end{equation}
It is easy to see that $E\in\mathbb{R}$ simply by acting with $(\psi,\cdot)$ and $(\cdot, \psi)$. One can also show that $E\in\mathbb{R}_+$. Namely, act with $(\psi,\cdot)$ on \eqref{schh} and explicitly write the inner product and use partial integration to get 
\begin{equation}
E=\left(\frac{\left\|\psi'\right\|}{\left\|\psi\right\|}\right)^2 > 0, \quad \forall \psi\in D(\hat{H}).
\end{equation}
Therefore we can define $E:=p^2$, $p\in\mathbb{R}$ and use the ansatz $\psi(x)\propto \text{e}^{ipx}$ to solve \eqref{schh}. After imposing boundary conditions and normalization condition, the solutions are given by
\begin{equation}\label{specc}
\psi_n=\frac{\text{e}^{ip_n x}}{\sqrt{L}}, \quad p_n=\frac{2n\pi}{L}, \quad E_n=p^{2}_{n}=\frac{4n^2 \pi^2}{L^2}, \quad n\in\mathbb{Z}.
\end{equation}
We see that the operators $\hat{H}$ and $\hat{p}$ have common eigenvectors. Notice an important property of the spectrum of the Hamiltonian, it is degenerate because the states $\psi_{\left|n\right|}$ and $\psi_{-\left|n\right|}$ have the same energy $E_n\propto n^2$. Degenerate spectrum is a signature of some symmetry in the system. Here the symmetry is generated by the momentum operator \eqref{mSA}, and since the Hamiltonian and $\hat{p}$ are well defined on the same domain, the formal commutator makes sense\footnote{Here the maximal invariant subspace of the algebra generated by $\hat{H}$ and $\hat{p}$ is $\mathcal{T}=\bigcap^{\infty}_{n,m=0}D(\hat{H}^n \hat{p}^m)=D(\hat{H})=D(\hat{p})$. }
\begin{equation}\label{comhp}
[\hat{H},\hat{p}]=0, \quad \psi=D(\hat{H}).
\end{equation}
We conclude that because we are dealing with properly defined SA operators $\hat{H}$ and $\hat{p}$ on the same domain $D(\hat{H})=D(\hat{p})$ rendering a well defined commutator \eqref{comhp}, we find no anomalous contribution \eqref{ano} and this symmetry is reflected in the degeneracy of the spectrum of the Hamiltonian. There is a physical interpretation of this symmetry if we let the circle be embedded in the $xy$-plane. The symmetry connected with the energy spectrum degeneracy is the rotation invariance around the $z$-axis, since the rotation generator $L_z=\hat{p}$ commutes with the Hamiltonian (for another analysis see \cite{Thiring}).\\

This conclusion will not be valid  if we switch from considering a particle on a circle (periodic boundary condition) to the finite interval (non-periodic conditions). We have already seen in subsections III.A and III.B that the momentum operator on a finite interval is a tricky business and using the von Neumann's method we found that there is a one-parameter family of non-equivalent  SA momentum operators given by
\begin{equation}\begin{split}\label{pf}
\hat{p}&=-i\frac{\d}{\d x}, \quad D_{\theta}(\hat{p})=\left\{\psi\left|\psi,\psi^{'}\in L^2[0,L] \ \text{and absolutely continuous and}, \psi(L)=\text{e}^{i\theta}\psi(0), \theta\in\mathbb{R}\ \text{mod}\  2\pi\right.\right\} \\
\hat{p}\psi^{\theta}_{n}&=p_n \psi^{\theta}_n, \quad \Longrightarrow \quad \psi^{\theta}_n=\frac{\text{e}^{ip_n x}}{\sqrt{L}}, \quad p_n=\frac{2n\pi +\theta}{L}, \quad n\in\mathbb{Z}.
\end{split}\end{equation}
The set $\left\{\psi^{\theta}_n\right\}^{+\infty}_{-\infty}$ forms an orthonormal basis in $L^2[0,L]$.\\

Now we turn to the Hamiltonian. This problem, the so called particle in a infinite square-well potential is worked out in all basic textbooks on QM. The only thing missing there are the details about the domain. So we fill this gap. We immediately define a proper domain, namely, the SA Hamiltonian is defined by
\begin{equation}\label{Hf}
\hat{H}=-\frac{\d^2}{\d x^2}, \quad D(H)=\left\{\psi\left| \psi,\psi^{'},\psi^{''}\in L^2[0,L]\ \text{and}\ \psi(0)=\psi(L)=0\right.\right\}.
\end{equation}
On such domain our Hamiltonian is SA and one can easily check that
\begin{equation}
(\psi,\hat{H}\varphi)=(\hat{H}\psi,\varphi), \quad \forall\psi,\varphi\in D(H) \quad \text{and} \quad D(\hat{H}^\dagger)=D(\hat{H}).
\end{equation}
We now turn to solving the eigenvalue equation
\begin{equation}
\hat{H}\psi=E\psi, \quad \psi\in D(\hat{H})
\end{equation}
and by the same arguments as for the particle on circle one can see that $E\in\mathbb{R}_+$, the normalized solutions are given by
\begin{equation}\label{solH}
\psi_n=\sqrt{\frac{2}{L}}\sin\left(\frac{n\pi x}{L}\right), \quad E_n=\frac{n^2\pi^2}{L^2}, \quad n\in\mathbb{N},
\end{equation}
and $\left\{\psi_n\right\}^{+\infty}_{1}$ form an orthonormal basis in $L^2 [0,L]$. It is important to note a qualitative difference between this spectrum and the spectrum obtained for the Hamiltonian of the particle on a circle \eqref{specc}. Namely, apart form the $2n\rightarrow n$ difference, there is a crucial difference in the ranges of $n$. For the particle on a circle we have $n\in\mathbb{Z}$, while for the finite interval $n\in\mathbb{N}$. Therefore there is no degeneracy in the spectrum of the Hamiltonian for the infinite square-well because there is a one-to-one correspondence  between each state $\psi_n$ and the value of its energy $E_n$. This is a consequence of the failure of the translation symmetry in this case. Namely, a SA momentum operator does not exists for the domain $D(\hat{H})$ so the use of the naive commutator
\begin{equation}\label{hpcom}
[\hat{H},\hat{p}]=0
\end{equation}
is meaningless. A close inspection shows that the maximal invariant subspace of the algebra generated by $\hat{H}$ defined by \eqref{Hf} and $\hat{p}$ defined by \eqref{pf} is 
\begin{equation}
\mathcal{T}=\bigcap^{\infty}_{n,m=0}D(\hat{H}^n \hat{p}^m)=\left\{\emptyset\right\},
\end{equation}
so there is no domain on which the commutator \eqref{hpcom} could even be defined, provided that the Hamiltonian and momentum are SA operators on a finite interval, and therefore there is no translation symmetry and no degeneracy of the spectrum of the Hamiltonian.\\

However, the measurement of both energy and momentum makes sense for both systems. Namely, one can calculate the momentum of each energy eigenstate, and the energy of each momentum eigenstate. For example, in the case of the particle on a circle, the states $\psi_n=\frac{1}{\sqrt{L}}\text{e}^{i\frac{2n\pi}{L}}$ are common eigevectors of both $\hat{p}$ and $\hat{H}$ so there is a sharp value of the momentum $p_n=\frac{2n\pi}{L}$ and energy $E_n=p^2_n$ for a given state $\psi_n$. In the case of a particle in a square-well potential, the energy eigenstates are given by $\psi_n=\sqrt{\frac{2}{L}}\sin\left(\frac{n\pi x}{L}\right)$ and are not in the domain of the SA momentum operator \eqref{pf}, but since $\psi_n\in L^2[0,L]$ it can be expanded in terms of the eigenvectors $\psi^{\theta}_{n}$ of the momentum operator 
\begin{equation}
\psi_n=\sum^{+\infty}_{m=-\infty}a^n_m\psi^{\theta}_{m}, \quad a^n_m:=(\psi^{\theta}_{m}, \psi_n)=\frac{\sqrt{2}}{\pi}\text{e}^{-i\frac{\theta}{L}}\frac{n}{n^2-4m^2}\left(1-(-)^n\right),
\end{equation}
so that the measurement of the momentum is not sharp since
\begin{equation}
\hat{p}\psi_n=\sum^{+\infty}_{m=-\infty}a^n_m\hat{p}\psi^{\theta}_{m}=\sum^{+\infty}_{m=-\infty}a^n_m \frac{2m\pi+\theta}{L}\psi^{\theta}_{m}\neq \alpha_n \psi_n,
\end{equation}
but rather there is a probability  $\left|a^n_m\right|^2$ to measure the value $p_m=\frac{2m\pi+\theta}{L}$. Similarly, the energy of the eigenstate of the momentum $\psi^\theta_n$ is not a sharp value, but rather there is a probability  $\left|a^m_n\right|^2$ to measure the energy $E_m=\frac{m^2\pi^2}{L^2}$.\\

Notice that we could arrive at the conclusion of the failure of the translation symmetry after noticing that the momentum operator on the class of functions which vanish at the end of the interval is only symmetric (see section III.A.) and not SA, and therefore it is impossible to construct a unitary representation of the translation. This means that such translations would  affect the normalization of states (they will not preserve the norm of the states), apart from not leaving the domain of the Hamiltonian invariant.\\

\subsection{Pauli's theorem}
Pauli's theorem states that time $t$ in QM can not be a SA operator conjugate to energy $E$, that is, it has to be regarded as an ordinary number, i.e. real parameter. The proof is an analog of the statement that momentum can not be a SA operator on a half line. If we demand that the time operator $\hat{t}$ is conjugate to energy operator $\hat{E}$, then we impose the canonical commutation relation
\begin{equation}\label{te}
[\hat{t},\hat{E}]=i.
\end{equation}
We can represent the operators $\hat{t}$ and $\hat{E}$ as 
\begin{equation}
\hat{t}:= i\frac{\d}{\d E}, \quad \hat{E}:= E.
\end{equation}
Since the energy of all physical systems has to be bounded from below, that is $E\in\left[E_{0},+\infty\right)$, we see that the deficiency indices of $\hat{t}$ are not equal,\footnote{Similarly as for the momentum operator on a half plane.} which implies that time cannot be realized as a SA operator conjugated to energy for systems whose energy is bounded from below.\\

The earliest proof, to the best of our knowledge, was given by W. Pauli \cite{WP}. He stated that since the eigenvalues for time range in $t\in\left( -\infty,+\infty\right)$, time can not be represented as a hermitian operator (depending only on phase space variables $x$ and $p$) and satisfy the commutation relation \eqref{te} since in that case energy should also range in $E\in\left( -\infty,+\infty\right)$, which contradicts the experience, that is the existence of a ground state (and point spectrum in general). He concludes that time has to be an ordinary number, a parameter in the theory.\\

However, for alternative discussion on how to avoid the Pauli's theorem see \cite{alterpauli}.

\section{Quantum mechanics in dimensions higher than 1}
 In this section we want to point out another subtle problem of defining observables that occurs when going to dimension higher than 1. So far we have seen that in order to have well defined  SA operators one needs to take care of domains. We will see that in order to pass from one dimension to two, one needs even more than just functional analysis to deal with SA, one actually needs the full machinery of differential geometry and fiber bundles \cite{Marsh:2016hdj}. However, since this goes beyond the scope of this paper I will just illustrate the problem and motivate the geometrical solution.\\

To this purpose let us consider QM in $d\ge 1$ dimensions. First let us define our wavefunctions as elements of a Hilbert space $\psi\in L^2(\mathbb{R}^d)$ where 
\begin{equation}
L^2=\mathcal{L}^2/\sim =\left\{[\psi] | \psi\in \mathcal{L}^2 \right\}
\end{equation}
and 
\begin{equation}
\mathcal{L}^2=\left\{ \psi:\mathbb{R}^d\rightarrow\mathbb{C}|\ \Re(\psi)\ \text{and}\ \Im(\psi) \ \text{are measurable and} \int_{\mathbb{R}^d}\left|\psi\right|^2d^d x<\infty \right\} 
\end{equation} 
with the inner product $(\cdot, \cdot): L^2 \times L^2 \rightarrow\mathbb{C}$ and $([\psi],[\varphi])\mapsto\int_{\mathbb{R}^d}\bar{\psi}\varphi d^dx$ that induces the equivalence relation $\psi\sim \varphi \Leftrightarrow \left\|\psi\right\|=\left\|\varphi\right\|$ and the integral is Lebesgue. In short, wavefunctions are square-integrable complex-valued functions. We will later see that exactly this has to be modified, and wavefunctions are actually sections on some appropriate bundle, but we will come to that later. Now we want to have our canonical commutation relation for coordinates $q^i$ and momenta $p_j$
\begin{equation}\label{dcr}
[q^i,p_j]=i\delta^i_j, \quad [q^i,q^j]=[p_i,p_j]=0, \quad i,j=\left\{1,...,d\right\}
\end{equation}
and realize them as SA operators $q^i, p_j : L^2\longrightarrow L^2$. We can represent them in the coordinate representation as multiplication operator and partial derivative respectively
\begin{equation}
(q^i\psi)(x):=x^i\psi(x), \quad (p_i\psi)(x):=-i\partial_i\psi(x)
\end{equation}
and immediately we encounter various domain and target issues because if $\psi\in L^2(\mathbb{R}^d)$, in general, such an element $\psi$ will either not lie in the SA domain of the operators, or the target will leave the space $L^2(\mathbb{R}^d)$. This problem can be solved, as discussed in previous sections, by suitable choice of domain. Here one can construct the maximally invariant subspace $\mathcal{T}:=\bigcap^{\infty}_{n,m=0}D(q^n p^m)$ which is exactly the Schwartz space\footnote{see Appendix G.} $\mathcal{S}(\mathbb{R}^d)$ and this resolves the domain and target issues since now $q^i, p_j : \mathcal{S}(\mathbb{R}^d)\longrightarrow \mathcal{S}(\mathbb{R}^d)$ renders the operators in question essentially SA.\\

The new kind of problem (that goes beyond the scope of functional analysis!) occurs if we want to switch from Cartesian to some other coordinate system. For example, let us take $d=2$ and try to do QM in polar coordinates. Therefore we make a coordinate transformation\footnote{For the inner product for $d=2$ we have $(\psi,\varphi)=\int dxdy \ \bar{\psi}(x,y)\varphi(x,y)=\int rdrd\phi\ \bar{\psi}(r,\phi)\varphi(r,\phi)$. }
\begin{equation}
x^i\longrightarrow r, \phi, \quad p_i\longrightarrow p_r=-i\partial_r, p_\phi=-i\partial_\phi
\end{equation}
and the commutation relations \eqref{dcr} still hold but we have a problem with self-adjointness (even worse, the operator $p_r$ is not even symmetric!). It is easy to see that $x^{i}=\left\{r,\phi\right\}$ are symmetric, and the same hold for $p_\phi$, $(\psi,p_\phi \varphi)=(p_{\phi}\psi,\varphi)$. But for the $p_r=-i\partial_r$ we have
\begin{equation}
(\psi,p_r\varphi)=-i\int rdrd\phi\ \bar{\psi}(r,\phi)\partial_r\varphi(r,\phi)=(p_r\psi,\varphi)+i\int drd\phi\ \bar{\psi}(r,\phi)\varphi(r,\phi)
\end{equation}
and the last term destroys the symmetry property. Therefore the quantization in coordinates other than Cartesian is not working at all. Notice that the switch from the Cartesian coordinates to the polar coordinates ``hides'' the space translation symmetry since it fixes the origin as a privileged point, and then, it is not so surprising that $p_r=-i\partial_r$ cannot be a SA generator of the space translations. \\

The solution to this problem is to realize that wave functions $\psi$ are not $\mathbb{C}$-valued functions, but rather  sections of a complex line bundle over\footnote{Or over some manifold $M$ in general.} $\mathbb{R}^d$, that is $\psi\in\Gamma(E)$ and $E$ is an appropriate line bundle $E\xrightarrow{\pi_E}\mathbb{R}^d$. If so, we could improve our quantization procedure by replacing the partial derivatives $\partial_i$ with covariant derivatives $\nabla_i$, knowing that covariant derivative acts differently on different objects, namely
\begin{equation}\begin{split}
\nabla_i f&=\partial_i f \ \text{if} \ f:\mathbb{R}^d\rightarrow\mathbb{C}\\
\nabla_i \gamma&=\partial_i\gamma+\omega_i\gamma \ \text{if} \ \gamma\in\Gamma(E)
\end{split}
\end{equation}
where $\omega_i$ is the Yang-Mills field. Therefore, let us propose that $\psi\in\Gamma(E)$ and $p_i=-i\nabla_i$. For the Cartesian coordinates we have $\omega_i=0$ and everything is ok. In the polar coordinates we had no problems with $p_\phi$ so $\omega_\phi=0$. For $p_r=-i\nabla_r$ in order to have the symmetry property $(\psi,p_r\varphi)=(p_r\psi,\varphi)$ we get a condition on $\omega_r$ as
\begin{equation}\label{conr}
2\Re(\omega_r)=\frac{1}{r}
\end{equation}
and it is easy to see that the canonical commutation relations still hold. Also, it is important to note that this construction works for any dimension $d$, and even if we move from flat $\mathbb{R}^d$ space to some curved space, i.e. Riemann manifold $(M,g)$. In that case, the analog analysis generalizes \eqref{conr} to
\begin{equation}
2\Re(\omega_j)=\partial_j (\log{\sqrt{g}}),
\end{equation} 
where $g$ is the determinate of the metric.\\

This all is justified by the fact that if $\psi$ is a section of a complex line bundle over $\mathbb{R}^d$, it locally looks like a $\mathbb{C}$-valued function. $E\xrightarrow{\pi_E}\mathbb{R}^d$ is an associated bundle to the frame bundle $LM\xrightarrow{\pi}\mathbb{R}^d$ and on a frame bundle we can establish a connection that gives rise to a covariant derivative. Also, sections of an associated bundles can be represented as $\mathbb{C}$-valued functions on the total space of the principle bundle $\psi:LM\longrightarrow\mathbb{C}$, and on the principle bundle exists an exterior covariant derivative $D\psi=d\psi+\omega\psi$. The theory of fiber bundles is one of the most important achievements in modern mathematics and it has numerous applications in various branches of mathematics and theoretical physics \cite{Balachandran:2017jha}, but its immediate occurrence in QM is quite remarkable \cite{Bohm:1993dr, knjigabohm, barry, senbundle}. Here we stop  any further discussion and end with the conclusion that requiring self-adjointness in higher dimensions and/or curved spaces leads to a natural appearance of the geometry of fiber bundles and QM seems to be fundamentally a gauge theory.

\section{Final Remarks}

In this paper we have analyzed and discussed some of the subtleties of representing observables in QM as SA operators. The key thing was to realize that operators in Hilbert space must be defined on some domain and choosing or finding a domain that renders the operator in question SA is necessary not only for performing practical calculations, but also to avoid some apparent paradoxes or some conceptual issues. Most of these issues steam either from  an ill-defined or under-defined operator as being hermitian (avoiding or suppressing the domains), or from solving the Schr\"{o}dinger equation and dealing with the Dirac ``bra-ket'' notation when it comes to examining the non-square integrable scattering states and continuous part of the spectrum. \\

In order to avoid or better to say pin-point some of these issues we suggest that only few new ingredients are necessary to be implemented in the ``standard'' exposition of the theory of QM.\\

 The first is, of course, the existence of domain of an operator. With that said  one can neatly refine the hermiticity in physics as \cite{morreti}:\\

\textit{A linear operator $T: D(T)\longrightarrow \mathcal{H}$ is 
\begin{enumerate}[(i)]
\item \textbf{hermitian} if $(T\psi,\varphi)=(\psi,T\varphi)$ $\forall \psi, \varphi\in D(T)$;
\item \textbf{symmetric} if it is densely defined ($\overline{D(T)}=\mathcal{H}$) and hermitian\footnote{This is equivalent to saying that the adjoint $T^\dagger$ is the extension of $T$, i.e. $T\subset T^\dagger$, meaning $D(T)\subset D(T^\dagger)$ and $T^\dagger \psi=T\psi$ $\forall\psi\in D(T)$. }; 
\item \textbf{self-adjoint} if it is symmetric and $T=T^\dagger$;
\item \textbf{essentially self-adjoint} if it is symmetric and $(T^\dagger)^\dagger=T^\dagger$.
\end{enumerate}}
Furthermore, it is important to emphasize that only (essentially)\footnote{Essentially self-adjoint operators have a unique SA exstension.} SA operators are the good representatives of observables in physics.\\

Secondly, one needs to better define the spectrum of an operator. For this purpose it is convenient to generalize the notion of eigenvalues. Remember that eigenvalues of a linear operator $T: D(T)\longrightarrow \mathcal{H}$ are some numbers $\lambda\in\mathbb{C}$ determined by the eigenvalue equation $T\psi=\lambda\psi$, $\forall\psi\in D(T)$. This is equivalent to stating that $\lambda$'s are such that the operator $(T-\lambda I)^{-1}$ does not exist\footnote{Notice that we assume that if $T: D(T)\longrightarrow \mathcal{H}$ is injective then $T^{-1}$ is the inverse restricted to the image of $T$, i.e. $T^{-1}:Im(T)\longrightarrow D(T)$.}. In finite dimensional Hilbert space the above statements are trivial, but in an infinite-dimensional Hilbert space some topological issues arise, namely, even if $(T-\lambda I)^{-1}$ exists it could be a bounded or an unbunded operator, also its domain $Im(T-\lambda I)$ can be dense or not. These are exactly the features that are suitable in order to generalize the notion of eigenvalues and differentiate between different types of spectra. For that purpose let us first define the so called \textit{resolvent set} $\rho(T)$ as:
\begin{equation}
\rho(T)=\left\{\lambda\in\mathbb{C} \ | \ (T-\lambda I) \ \text{is injective,} \ \overline{Im(T-\lambda I)}=\mathcal{H} \ \text{and}\ \ (T-\lambda I)^{-1} \ \text{is bounded}   \right\}.
\end{equation}
Then\\

\textit{the \textbf{spectrum} of $T$ is the complement of the resolvent set, i.e. $\sigma(T):=\mathbb{C}-\rho(T)$, and is given by the union of (pairwise disjoint) parts:
\begin{enumerate}[(i)]
\item \textbf{point spectrum} $\sigma_{p}(T)$, where $(T-\lambda I)$ is not injective so that $\sigma_{P}(T)$ is actually the set of all the eigenvalues of $T$;
\item \textbf{continuous spectrum} $\sigma_{c}(T)$, where $(T-\lambda I)$ is injective, $\overline{Im(T-\lambda I)}=\mathcal{H}$ and $(T-\lambda I)^{-1}$ is not bounded; 
\item \textbf{residual spectrum} $\sigma_{r}(T)$, where $(T-\lambda I)$ is injective, $\overline{Im(T-\lambda I)}\neq\mathcal{H}$.
\end{enumerate}
}
It immediately follows that $\rho(T)$ is always an open set, meaning that the spectrum $\sigma(T)$ is always a closed set, and that for a SA operator $\sigma(T)\in\mathbb{R}$ and $\sigma_r(T)=\emptyset$ \cite{morreti}. So one could, in principle, avoid the unfortunate use of the non-square integrable solutions of the Schrodinger equation when dealing with the continuous spectrum by working with only a well defined object, the so called \textbf{resolvent} $R_{T}(\lambda)=(T-\lambda I)^{-1}$ and this way ``stay'' within the Hilbert space $\mathcal{H}$. Notice that the Green function is the integral kernel of the resolvent and solving the differential equation for the Green function makes perfect sense \cite{Madrid, madrid}.\\

The advantages and pitfalls of the Dirac's ``bra-ket'' notation are extensively and pedagogically discussed in \cite{Gieres:1999zv} and if one really wants to ``save it''  one needs to resort not only to the functional analytic and spectral theory tools but also to a wast machinery of the distribution theory as well. This all culminates into defining the so called rigged Hilbert space or Gelfand triple \cite{rigged}. But the point is that this could be avoided considering just the resolvent or the so called spectral theorem\footnote{For the details on spectral theorems for unbounded SA operators see for example \cite{r-s-1, glaz, morreti}. In order to properly understand these spectral theorems one needs a bit more then just a formal introduction to functional analysis. }, which are the rigorous generalization and justification of the formal manipulation within the Dirac's formalism, but without ever ``leaving'' the Hilbert space.

\bigskip
\bigskip

\noindent{\bf Acknowledgment}\\
 I  would like to thank Kumar S. Gupta for introducing me to this subject. The idea and much of the discussions here is based on or inspired by lectures and discussion with Kumar S. Gupta. Also, thanks to Prof. K. Veselic for his lectures, advices and giving me a chance to investigate this subject during his course long time ago.  Great thanks to my colleagues I.Smoli\'c and H.Nikoli\'c for numerous discussions, comments and for reading the manuscript. I am also thankful for numerous discussions with N. Kova\v{c}i\'c. I would like to thank B. Klajn for various discussions in the early stages of this work. Finally, I would like to thank the anonymous referees on their valuable comments and suggestions. This research has been supported by Croatian Science Foundation project IP-2020-02-9614.

\bigskip
\bigskip

\appendix
\section{Postulates of QM}
Here we will briefly outline the postulates of the so called \textsl{canonical quantization}\footnote{There are mainly three well established formulation of QM: 1. Heisenberg's matrix mechanics (historically first, but of less practicable use for a working physicist); Scr\"{o}dinger's wave mechanics (standard and most used formalism); Feynman's path integral (wildly used in relativistic quantum field theories). Actually, in general, one can formulated consistent non-relativistic QM in (at least) nine independent ways \cite{qmfor}}. In general, quantization is a procedure of constructing a quantum theory (QT) for a given classical system. This procedure is by no means unique and a rigorous approach to it is still not fully developed \cite{quantization}. There is only two things we require from such a procedure. One is that the QT must coincide with the predictions of the classical theory in the formal $\hbar\rightarrow 0$ limit\footnote{In mathematics, quantization is a quantum deformation of classical structures; the deformation parameter is the Planck constant $\hbar$.}(actually we need a bit more than that, like for an example, the limit to macroscopic scale(large masses), smooth potentials,  large quantum numbers, etc.). This is the essence of the ``correspondence principle''. The second requirement is that the prediction of the QT are in agreement with the experiments. Experience in quantization of simplest systems (such as a free particle, a harmonic oscillator, and a non-relativistic particle in some potential fields) was used to formulate a consistent general scheme of operator quantization for an arbitrary system with canonical Hamiltonian equations of motions for phase-space variables. This scheme is called \textsl{canonical quantization} and its \textsl{postulates} are:

\begin{enumerate}

\item For a given physical system we assume the existence of a Hamiltonian formulation of classical mechanics. States of the system are points in an even-dimensional phase space $\mathcal{P}=T^{*}M$  and are labeled by canonical generalized coordinates $q_{a}$ and momenta $p_{a}$, $a=1,...,n$, where $n$ is the number of degrees of freedom. Hamilton equations of motion govern the time evolution of the system
\begin{equation}
\dot{q}_{a}=\left\{q_{a},H\right\}, \quad \dot{p}_{a}=\left\{p_{a},H\right\},
\end{equation}
where $H=H(q,p)$ is the Hamiltonian of the system and $\left\{,\right\}$ is the canonical Poisson bracket. The Poisson bracket of two arbitrary functions $f$ and $g$ on the phase space is given by
\begin{equation}
\left\{f,g\right\}=\sum_{a}\left(\frac{\partial f}{\partial q_{a}}\frac{\partial g}{\partial p_{a}}-\frac{\partial f}{\partial p_{a}}\frac{\partial g}{\partial q_{a}}\right),
\end{equation}
and in particular, we have $\left\{q_{a},q_{b}\right\}=\left\{p_{a},p_{b}\right\}=0$ and $\left\{q_{a},p_{b}\right\}=\delta_{ab}$. Classical observables $f:\mathcal{P}=T^{*}M\longrightarrow\mathbb{R}$ are local physical quantities described by real functions of the phase-space variables and they form a real associative commutative algebra.

\item A state in QM is defined as a vector $\psi$ in a suitable Hilbert space $\mathcal{H}$. A scalar product of such two vectors $\psi_{1}$ and $\psi_{2}$ is denoted by $(\psi_{1},\psi_{2})$. It is assumed that any state $\psi\in\mathcal{H}$ can be realized physically (or at least to a first approximation) and that the superposition principle holds, that is if states $\psi_{1}$ and $\psi_{2}$ are realizable, then the state $\psi=a_{1}\psi_{1}+a_{2}\psi_{2}$ with any $a_{1},a_{2}\in\mathbb{C}$ is also realizable.

\item To an each classical observable $f=f(x,p)$ we uniquely assigned a linear SA operator $\hat{f}$ acting in Hilbert space $\mathcal{H}$.  The operator $\hat{f}$ is called a quantum observable. It is assumed that any operator $\hat{f}$ is well defined\footnote{To be precise this only holds for finite-dimensional Hilbert spaces and  for bounded operators in infinite-dimensional Hilbert spaces. In physics we are usually dealing with unbounded operators, for which every operator has its domain.} on any state $\psi$, i.e., $\hat{f}\psi\in\mathcal{H}$, $\forall\psi\in\mathcal{H}$. If so, the operator $\hat{f}$ is  uniquely determined by its matrix elements $(\psi_{1},\hat{f}\psi_{2})$, $\forall\psi_{1}\psi_{2}\in\mathcal{H}$, that is by its matrix $f_{mn}=(e_{m},\hat{f}e_{n})$ with respect to an orthonormal basis\footnote{$\left\{\psi_{n}\right\}^{\infty}_{1}$ is an infinite sequence of vectors.} $\left\{e_{n}\right\}^{\infty}_{1}$ in $\mathcal{H}$. To any  such operator $\hat{f}$ we can assign its adjoint $\hat{f}^\dagger$ 
\begin{equation}
\left(\psi_{1},\hat{f}^\dagger\psi_{2}\right)=\left(\hat{f}^\dagger\psi_{1},\psi_{2}\right), \quad \forall\psi_{1},\psi_{2}\in\mathcal{H},
\end{equation}
and thereby the involution (conjugation) $\hat{f}\mapsto\hat{f}^\dagger$ is defined in the algebra of operators with the properties
\begin{equation}\begin{split}
&\left(\hat{f}^{\dagger}\right)^{\dagger}=\hat{f}, \quad \left(a\hat{f}\right)^{\dagger}=a^*\hat{f}^{\dagger}, \quad \forall a\in\mathbb{C}\\
&\left(\hat{f}+\hat{g}\right)^{\dagger}=\hat{f}^{\dagger}+\hat{g}^{\dagger}, \quad \left(\hat{f}\hat{g}\right)^{\dagger}=\hat{g}^{\dagger}\hat{f}^{\dagger}.
\end{split}\end{equation}
The operator $\hat{f}$ is SA\footnote{More precisely see Appendix C.} if $\hat{f}=\hat{f}^{\dagger}$, or
\begin{equation}
\left(\psi_{1},\hat{f}\psi_{2}\right)=\left(\hat{f}\psi_{1},\psi_{2}\right), \quad \forall\psi_{1},\psi_{2}\in\mathcal{H}
\end{equation}
The expectation value $\left\langle \hat{f}\right\rangle_{\psi}$ of any quantum observable $\hat{f}$ in a state $\psi$ and the corresponding dispersion $\Delta f$ are defined as
\begin{equation}
\left\langle \hat{f}\right\rangle_{\psi}=\frac{\left(\psi,\hat{f}\psi\right)}{\left(\psi,\psi\right)}, \quad \Delta \hat{f}=\sqrt{\left\langle \hat{f}^2\right\rangle_{\psi}-\left\langle \hat{f}\right\rangle^{2}_{\psi}}.
\end{equation}
The observables are assumed to be SA operators because the corresponding eigenvalues are real and the eigenvectors form an orthonormal basis in $\mathcal{H}$. The spectrum (set of all the eigenvalues) represent all the possible measurements, while the complete orthonormalized set of the eigenstates of the observable provides a probabilistic interpretation of its measurements.

\item The correspondence principle implies a connection between the Poisson bracket of classical observables and the commutator of the quantum observables. Namely,
\begin{equation}
\left\{f_{1},f_{2}\right\}\rightarrow\frac{1}{i\hbar}\left[\hat{f}_{1},\hat{f}_{2}\right]+\hat{O}(\hbar)
\end{equation}
The position operators $\hat{q}_{a}$ and momentum operators $\hat{p}_{a}$ are postulated to be SA and satisfy the canonical commutation relations
\begin{equation}
\left[\hat{q}_{a},\hat{q}_{b}\right]=\left[\hat{p}_{a},\hat{p}_{b}\right]=0, \quad \left[\hat{q}_{a},p_{b}\right]=i\hbar\left\{q_{a},p_{b}\right\}=i\hbar\delta_{ab}
\end{equation}

The correspondence principle imposes the form of the quantum observable as $\hat{f}=f(\hat{x},\hat{p})+\hat{O}(\hbar)$, where $\hat{O}(\hbar)$ is chosen in such a way that it insures the self-adjointness. Since $\hat{q}_a$ and $\hat{p}_a$ don’t commute, due to the so-called ordering problem\footnote{A substantial contribution to the resolution of this problem is due to Berezin \cite{berezin}}, there is no unique construction of $f(\hat{x},\hat{p})$ via 
$f(x,p)$. Any two commuting observables $\hat{f}_{1}$ and $\hat{f}_{2}$ can be simultaneously measured because the commutativity implies the existence of common eigenvectors and joint spectrum. A minimum set of $N$ commuting observables $\hat{f}_{k}$, $k=1,...,N,$ $[\hat{f}_{k}, \hat{f}_{l}]=0$, $\forall k, l$ whose joint spectrum is nondegenerate and whose common eigenvectors provide a unique specification of any vector in terms of the corresponding expansion with respect to these eigenvectors define a complete set of observables. Complete sets of observables completely specify the quantum description of a system under consideration.

\item The time evolution of any quantum state $\psi(t)$  is governed by the Schr{\"o}dinger equation, 
\begin{equation}
i\hbar\frac{\partial\psi}{\partial t}=\hat{H}\psi,
\end{equation}
with an initial condition $\psi(t_{0})=\psi_{0}$, where the operator $\hat{H}$ corresponds to the classical Hamiltonian $H$. 
\end{enumerate}

This concludes the postulates of canonical quantization which are usually presented in most standard textbooks on QM. When taken seriously, these postulates will lead to many paradoxes, some of which we discuss in section II.

\section{Bounded linear operators}
Let $(\mathcal{V},\left\|\cdot\right\|_{\mathcal{V}})$ be a normed space and $(\mathcal{W},\left\|\cdot\right\|_{\mathcal{W}})$ a Banach space\footnote{A Banach space is a normed and complete space (all Cauchy sequences are convergent in it).}. A linear map $T:\mathcal{V}\longrightarrow\mathcal{W}$ is called a bounded linear operator if
\begin{equation}
\underset{f\in\mathcal{V}\setminus\left\{0\right\}}{\text{sup}}\frac{\left\|Tf\right\|_{\mathcal{W}}}{\left\|f\right\|_{\mathcal{V}}} < \infty
\end{equation}
For a bounded operator one can define the so called operator norm $\left\|\cdot\right\|$ such that
\begin{equation}
\left\|T\right\|=\underset{f\in\mathcal{V}\setminus\left\{0\right\}}{\text{sup}}\frac{\left\|Tf\right\|_{\mathcal{W}}}{\left\|f\right\|_{\mathcal{V}}}
\end{equation}
and there are theorems that state that a linear map is continuous\footnote{with respect to the topologies induced by the respective
norms on $\mathcal{V}$ and $\mathcal{W}$} if and only if this map is a bounded operator and that a bounded operator is defined on all of $\mathcal{V}$, that is $D(T)=\mathcal{V}$. For these operators the naive quantization prescription would work much better, but unfortunately (or fortunately!) in physics we almost always have to deal with observables that are unbounded (the operator norm is ill-defined), like the paradoxes in section II illustrate. Also note that a Hilbert space is always a Banach space since the inner product $(\cdot, \cdot)_{\mathcal{H}}$ induces a norm $\left\|\cdot\right\|_{\mathcal{H}}=\sqrt{(\cdot, \cdot)_{\mathcal{H}}}$. Notice that the set of all continues linear mappings $\mathcal{L}:\mathcal{H}\longrightarrow\mathcal{H}$ form a Banach space (even more, a Banach algebra), which is the space of all the bounded operators $\mathcal{L}=\mathcal{B}(\mathcal{H})$ on $\mathcal{H}$.

\section{Hilbert spaces}
In this appendix, we outline the properties of  Hilbert spaces. The shortest way to define a Hilbert space $\mathcal{H}$ is that it is a Banach space with respect to a norm induced by the sesquilinear inner product $(\cdot, \cdot):\mathcal{H}\times\mathcal{H}\longrightarrow\mathbb{C}$. The norm is $\left\|\cdot\right\|=\sqrt{(\cdot, \cdot)}$. This is very concise so let us outline all the properties of Hilbert spaces in more details:
\begin{itemize}
\item Hilbert space $\mathcal{H}$ is a vector space over the complex numbers. Usually, the elements (also called vectors or points) of $\mathcal{H}$ we denote by Greek letters, while the complex numbers with Latin letters. The vector space structure is encoded in $(\mathcal{H}, \cdot, +)$ where $\cdot:\mathbb{C}\times\mathcal{H}\longrightarrow\mathcal{H}$ is the scalar multiplication and $+:\mathcal{H}\times\mathcal{H}\longrightarrow\mathcal{H}$ is the vector addition satisfying
\begin{equation}
a\psi+b\eta \in \mathcal{H}\  \forall a,b\in\mathbb{C}\  \text{and}\  \psi, \eta\in\mathcal{H}
\end{equation}
and the vector space axioms\footnote{For all $\psi,\eta, \xi\in\mathcal{H}$ and $a, b\in\mathbb{C}$ we have
\begin{equation}\begin{split}
&\text{(Commutativity of addition)} \ \psi+\eta=\eta+\psi\\
&\text{(Associativity of addition)} \ (\psi+\eta)+\xi=\psi+(\eta+\xi)\\
&\text{(Existence of a zero vector)} \ \text{There is a vector}\ \ o\in\mathcal{H}\ \text{with}\ \ o+\psi=\psi+o=\psi\\
&\text{(Existence of additive inverses)} \ \text{For each}\ \psi,\ \text{there is}\ -\psi\in\mathcal{H}\ \text{such that}\ \ \psi+(-\psi)=(-\psi)+\psi=o\\
&\text{(Distributivity of scalar multiplication over vector addition)}\ \ a(\psi+\eta)=a\psi+a\eta\\
&\text{(Distributivity of scalar addition over scalar multiplication)}\ \ (a+b)\psi=a\psi+b\psi\\
&\text{(Associativity of scalar multiplication)}\ \ (ab)\psi=a(b\psi)\\
&\text{(Scalar multiplication with 1 is the identity)}\ \ 1\psi=\psi\\
\end{split}\end{equation}}

\item The inner product $(\cdot, \cdot):\mathcal{H}\times\mathcal{H}\longrightarrow\mathbb{C}$ is a positive definite sesquilinear form on $\mathcal{H}$ satisfying
\begin{equation}\begin{split}\label{inner}
&(\xi,\eta)=\overline{(\eta,\xi)}; \quad (\xi,\xi)\geq 0, \quad \text{and} \quad (\xi,\xi)=0 \iff \xi=0;\\
&(\xi,a\zeta+b\eta)=a(\xi,\zeta)+b(\xi,\eta)\Rightarrow (a\xi+b\zeta,\eta)=\bar{a}(\xi,\eta)+\bar{b}(\zeta,\eta).
\end{split}\end{equation}
and induces a norm  defined by $||\xi||=\sqrt{(\xi,\xi)}$. The norm $\left\|\cdot\right\|:\mathcal{H}\longrightarrow\mathbb{R}_{+}$ satisfies the triangle inequality 
\begin{equation}\label{triangle}
||\xi+\eta||\leq||\xi||+||\eta||
\end{equation}
as a corollary of the \textsl{Cauchy-Schwarz-Bunyakovskii inequality} $|(\xi,\eta)|\leq||\xi||\;||\eta||$. The norm induces a metric $d(\xi, \eta):=\left\|\xi-\eta\right\|$ which determines the topology\footnote{A Hilbert space is a particular case of a normed and metric space in which a norm and a metric (distance) satisfying standard requirements are generated by a scalar product; see \cite{glaz}} in $\mathcal{H}$. Existance of topology gives meaning to convergence, so for a sequence of vectors $\left\{\xi_{n}\right\}^{\infty}_{1}$ we say that it converges to a vector $\xi$ if
\begin{equation}
\xi=\lim_{n\rightarrow\infty}\xi_{n}
\end{equation}
or equivalently $\lim_{n\rightarrow\infty}||\xi_{n}-\xi||=0$. Due to \eqref{triangle} the necessary condition for convergence is
\begin{equation}\label{conv}
\lim_{n\rightarrow\infty}||\xi_{m}-\xi_{n}||= 0, 
\end{equation}
i.e. the sequence $\left\{\xi_{n}\right\}^{\infty}_{1}$ is a Cauchy sequence. Linear operations in $\mathcal{H}$ (like multiplication of vectors by complex numbers and vector addition) and the scalar product are continuous in their arguments; for example, 
\begin{equation}
\lim_{n\rightarrow\infty}||\xi_{m}-\xi_{n}||= 0 \Rightarrow \lim_{n\rightarrow\infty}(\xi_{n},\eta)=(\xi,\eta), \quad \forall \eta\in\mathcal{H}.
\end{equation}
because of the Cauchy-Schwarz-Bunyakovskii inequality. A set $M\subset\mathcal{H}$ is  dense in $\mathcal{H}$ if any vector in $\mathcal{H}$ can be approximated by vectors belonging to $M$ with any desired accuracy, that, if for any $\xi\in\mathcal{H}$, there exists a sequence $\left\{\xi_{n}\right\}^{\infty}_{1}$, $\xi_{n}\in M$, so that $\xi=\lim_{n\rightarrow\infty}\xi_{n}$. In other words, the topological closure of $M$, that is $M$ and all of its Cauchy sequences, is equal to $\mathcal{H}$, i.e. $\overline{M}=\mathcal{H}$.
\item $\mathcal{H}$ is a complete normed space. This means that every Cauchy sequence $\left\{\xi_{n}\right\}^{\infty}_{1}$ in $\mathcal{H}$ is convergent, i.e. it has a limit in $\mathcal{H}$:
\begin{equation}
\lim_{n\rightarrow\infty}||\xi_{m}-\xi_{n}||= 0 \quad\ \Longrightarrow\quad\ \exists\xi\in\mathcal{H}: \lim_{n\rightarrow\infty}\xi_{n}=\xi.
\end{equation}
Notice that any convergent sequence $\left\{\xi_{n}\right\}^{\infty}_{1}$ is a Cauchy sequence and in a Hilbert space, the converse also holds\footnote{In short, a Hilbert space is complete with respect to a metric generated by a scalar product.}. A pre-Hilbert space is a vector space with an inner product satisfying \eqref{inner}. Physicists usually deal with just a pre-Hilbert space, since there is no difference between pre-Hilbert space and a Hilbert space of finite dimension. Fortunately, any pre-Hilbert space can be made a complete Hilbert space by adding the ``limits'' of all Cauchy sequences. Note that the requirement of completeness is crucial, and not only technical, for applications of Hilbert spaces to QM.
\item A Hilbert space $\mathcal{H}$ is called separable if it contains a countable dense set. Separable Hilbert spaces are sufficient for treating conventional QM. All  infinite-dimensional separable
Hilbert spaces are isomorphic to the Hilbert space of complex square summable sequences $l^2(\mathbb{N})$ defined by
\begin{equation}
l^2(\mathbb{N}):=\left(\left\{a:\mathbb{N}\longrightarrow\mathbb{C}|\sum^{\infty}_{n=0}\left|a_n\right|^2<\infty\right\}, (a,b)_{l^2}:=\sum^{\infty}_{n=0}a^{*}_{n}b_{n}\right)
\end{equation}
\item The vector space of all square-integrable functions on a real interval $(a,b)$  is usually denoted by $L^2(a,b)$ and can be made into a Hilbert space
\begin{equation}
L^2(a,b)=\left\{\psi(x):\int^{b}_{a}\text{d}x\ |\psi(x)|^2<\infty\right\}.
\end{equation}
The scalar product in $L^2(a,b)$ is defined by
\begin{equation}
(\psi_{1},\psi_{2})=\int^{b}_{a}\text{d}x\ \psi^{*}_{1}(x)\psi_{2}(x).
\end{equation}
It is important to note that here the integrals are Lebesgue integrals, and strictly speaking, the elements of $L^2(a,b)$ are equivalence classes of functions that are equal almost everywhere\footnote{When speaking about some function belonging to $L^2(a,b)$ and possessing some additional specific properties like absolute continuity, we actually mean the representative of the corresponding equivalence class.}.
\end{itemize}

\section{Adjoint of an operator and its properties}
\textbf{Definition.} The adjoint $A^\dagger$ of a densely defined linear operator $A:D(A)\longrightarrow\mathcal{H}$ is defined by
\begin{enumerate}
\item $D(A^\dagger):=\left\{\psi\in\mathcal{H}|\exists\eta\in\mathcal{H}:\forall\alpha\in D(A):(\psi,A\alpha)=(\eta,\alpha)\right\}$
\item $A^\dagger \psi=\eta$
\end{enumerate}

\textbf{Definition.} A densely defined\footnote{Densely defined means that the domain $D_f$ is dense in $\mathcal{H}$, i.e. $\overline{D_f}=\mathcal{H}$} linear operator $A:D(A)\longrightarrow\mathcal{H}$ is called symmetric if $\forall\alpha, \beta\in D(A): (\alpha, A\beta)=(A\alpha,\beta)$.\\

\textbf{Lemma.} If $A$ is symmetric, then $A\subseteq A^\dagger$ meaning
\begin{enumerate}
\item $D(A)\subseteq D(A^\dagger)$
\item $A^\dagger \psi=A\psi,\  \  \forall \psi\in D(A)$
\end{enumerate}

\section{Self-adjoint operators and their properties}
\textbf{Definition.} An operator $A:D_{A}\longrightarrow\mathcal{H}$ defined on a dense domain is called a \textsl{self-adjoint} operator (SA operator) if it coincides with its adjoint, $A=A^\dagger$. In the language of maps, this means that $A$ is symmetric, and $D_{A}=D_{A^\dagger}$.\\

\textbf{Lemma.} A \textsl{symmetric operator} $A$ is SA iff $\phi\in D_{A^\dagger}\Rightarrow\phi\in D_{f}$, i.e.,
\begin{equation}
\left(\phi,A\psi\right)=\left(A^\dagger \phi,\psi\right), \quad \forall \psi\in D_{A}\Rightarrow \phi\in D_{A}
\end{equation}
\\
To make sure that an operator $A$ is indeed SA, we must first verify that  $A$ is symmetric and then that the criterion of the above Lemma holds. Very often physicist just verify the symetricity property but omit to check the condition about the domain of the adjoint. However, observables must be represented by SA operators, because only they  have the remarkable property of having a real spectrum and a complete orthogonal system of (in general ``generalized'') eigenvectors corresponding to this spectrum, which provides the possibility of a probabilistic interpretation of states, observables, and measurements. Symmetric operators just have real expectation values.\\

\textbf{Lemma.} All expectation values of an SA operator $A$ are real, $(\psi,A\psi)=(\psi,A\psi)^*$, $\forall\psi\in D_{A}$, and define a norm of the operator,
\begin{equation}
\left\|A\right\|=\underset{\psi\in D_{A}, ||\psi||=1}{\text{sup}}\left|(\psi,A\psi)\right|.
\end{equation}

\textbf{Lemma.} SA operators have the following properties:
\begin{itemize}
\item For $A=A^{\dagger}$ and  $a\in\mathbb{R}\ \Rightarrow\  aA=(aA)^\dagger, D_{aA}=D_{A}$.
\item For $A=A^{\dagger}$ and  $B=B^{\dagger}$, with  $\overline{D_{A}\cap D_{B}}=\mathcal{H}\ \Rightarrow\  (A+B)^{\dagger}\supseteq A+B$.\\
The sum of two SA operators $A+B$ is  just a symmetric operator, in general, but if one of the operators, for example $B$, is  bounded, then the sum is also an SA operator, i.e. $A+B=(A+B)^\dagger$ with $D_{A+B}=D_{A}$.
\item For $A=A^{\dagger}$ and  $B=B^{\dagger}$, with $\overline{D_{AB}}=\mathcal{H}\ \Rightarrow (AB)^\dagger\supseteq BA$.\\
 The product $AB$ of two SA operators is not even symmetric. The product $AB$ is symmetric, i.e.  $(AB)^\dagger\supseteq AB$, if $A$ is bounded, and  $BA\subseteq AB$, that is if $A$ and $B$ commute. The product $AB$ is SA, i.e. $(AB)^\dagger=AB$, if both operators are bounded, and they commute  $[A,B]=0$.
\end{itemize}

These properties are important in constructing observables in QM. The second property is vital in constructing Hamiltonians. Namely, if we want to perturb the unbounded SA Hamiltonian of the free particle $\hat{H}_{0}$ by some bounded SA potential $\hat{V}$, then the total Hamiltonian $\hat{H}=\hat{H}_{0}+\hat{V}$ is indeed a SA operator on a domain $D_{\hat{H}}=D_{\hat{H}_{0}}$. However, if $\hat{V}$ is unbounded, then $\hat{H}$ is at most symmetric and we have to deal with the theory of SA extensions.\\

Notice that for general unbounded operators the notion of commutativity is ambiguous due to the domain issues. However, unbounded SA operators allow for the notion of commutativity in terms of corresponding one-parameter families $\left\{\hat{U}_{A}(\alpha)=\text{exp}(\text{i}\alpha A), \alpha\in\mathbb{R}\right\}$ of mutually commuting unitary operators\footnote{Note that unitary operators are bounded and defined everywhere, and the notion of commutativity for such operators is unambiguous.}
\begin{equation}
\left[\hat{U}_{A}(\alpha),\hat{U}_{A}(\beta)\right]=0, \forall\alpha, \beta \in \mathbb{R},
\end{equation}
associated with each SA operators $A$. We say that SA operators $A$ and $B$  commute, if the respective families $\left\{\hat{U}_{A}(\alpha)\right\}$ and $\left\{\hat{U}_{B}(\alpha)\right\}$ of the associated unitary operators mutually commute: $\left[\hat{U}_{A}(\alpha),\hat{U}_{B}(\beta)\right]=0, \forall\alpha, \beta \in \mathbb{R}$. 

For example, the canonical commutation relation $[\hat{q},\hat{p}]=i\hbar$ for the position operator $\hat{q}$ and the momentum operator $\hat{p}$ is properly formulated as the Weyl relation \cite{r-s-1}
\begin{equation}
\hat{U}_{q}(\alpha)\hat{U}_{p}(\beta)=\text{e}^{-\text{i}\alpha\beta\hbar}\hat{U}_{p}(\beta)\hat{U}_{q}(\alpha), \quad \forall\alpha, \beta \in \mathbb{R},
\end{equation}
for the corresponding associated unitary operators $\hat{U}_{q}(\alpha)=\text{exp}(\text{i}\alpha\hat{q})$ and $\hat{U}_{p}(\beta)=\text{exp}(\text{i}\beta\hat{p})$.

\section{The von-Neumann's theorems}
\textbf{The first von Neumann theorem}\\

For any symmetric operator $A:D_A\longrightarrow\mathcal{H}$ the domain of its adjoint $A^\dagger$ is given by the direct sum of subspaces \footnote{Where $\mathbb{C}^\prime$ is a set of complex numbers with nonzero imaginary part, $\mathbb{C}^\prime=\left\{z=x+iy, y\neq 0\right\}=\mathbb{C}_{+}\cup\mathbb{C}_{-}$. If we define $A(z)=A-zI$, then $\Sigma_{z}=\text{ker}A^\dagger(z^*)=\left\{\xi_{z^*}\in D_{A^\dagger}:A^\dagger \xi_{z^*}=z^* \xi_{z^*}\right\}$ and $\Sigma_{z^*}=\text{ker}A^\dagger(z)=\left\{\xi_{z}\in D_{A^\dagger}:A^\dagger \xi_{z}=z\xi_{z}\right\}$, and $\bar{A}$ is the closure of  $A$} $D_{\bar{A}}, \Sigma_{z^*}$ and $\Sigma_{z}$:
\begin{equation}
D_{A^\dagger}=D_{\bar{A}}+\Sigma_{z^*}+\Sigma_{z}, \quad \forall z\in\mathbb{C}^{\prime},
\end{equation}
such that any vector $\xi_{*}\in D_{A^\dagger}$ is uniquely represented as
\begin{equation}
\xi_{*}=\underline{\xi}+\xi_{z}+\xi_{z^*}, \quad \underline{\xi}\in D_{\bar{A}}, \quad \xi_{z}\in \Sigma_{z^*}, \quad \xi_{z^*}\in\Sigma_{z},
\end{equation}
which is known as the \textsl{first von Neumann formula} and
\begin{equation}
A^\dagger \xi_{*}=\overline{A}\underline{\xi}+z\xi_{z}+z^*\xi_{z^*}.\\
\end{equation}

\textbf{The second von Neumann theorem}\\

We say that a symmetric operator $A$ is essentially SA iff its deficiency indices\footnote{Deficiency index is defined as $m(z)=\text{dim}\ \text{ker} A^\dagger (z^*)=\left\{m_+, \ \text{for}\ \ z\in\mathbb{C}_{+}\ \text{or}\ \ m_{-} \text{for}\ \ z\in\mathbb{C}_{-}\right\}$} are equal to zero, $m_{\pm}=0$. A symmetric operator $\hat{f}$ is said to be essentially maximal, that is it does not allow SA extensions iff one of its deficiency indices is equal to zero,  while the other is nonzero. If both deficient subspaces $\Sigma_{z^*}$ and $\Sigma_{z}$ of a symmetric operator $A$ are nonzero, then ther exists a nontrivial symmetric extension of $A$. Symmetric extensions $A_{U}$ of a symmetric operator $A$ are determined by an isometric operator $U: D_{U}\subseteq\Sigma_{z^*}\longrightarrow UD_{U}\subseteq\Sigma_{z}$. These extensions are given by 
\begin{equation}\begin{split}\label{1}
D_{A_{U}}&=D_{\bar{A}}+(I+U)D_{U}\\
&=\left\{\xi_{U}: \xi_{U}=\underline{\xi}+\xi_{z,U}+U\xi_{z,U};\ \forall\underline{\xi}\in D_{\bar{A}},\forall\xi_{z,U}\in D_{U}\subseteq\Sigma_{z^*}; U\xi_{z,U}\in UD_{U}\subseteq\Sigma_{z}\right\},
\end{split}\end{equation}
and
\begin{equation}\label{2}
A_{U}\xi_{U}=\overline{A}\underline{\xi}+z\xi_{z,U}+z^*\hat{U}\xi_{z,U}.
\end{equation}
In other words, any isometric operator $U:\Sigma_{z^*}\rightarrow\Sigma_{z}$ with the domain $D_{U}\subseteq\Sigma_{z^*}$ and the range $UD_{U}\subseteq\Sigma_{z}$ defines a symmetric extensions $A_{U}$ of a symmetric operator $A$ given by \eqref{1} and \eqref{2}. The equality
\begin{equation}
\xi_{U}=\underline{\xi}+\xi_{z,U}+U\xi_{z,U}
\end{equation}
in \eqref{1} is usually called the \textsl{second von Neumann formula}.\\

\textbf{The main theorem}\\

Let $A:D_A\longrightarrow\mathcal{H}$ be a  symmetric operator  and  $A^\dagger$ its adjoint, i.e. $A\subseteq A^\dagger$. Let $\Sigma_{z^*}$ and $\Sigma_{z}$ be the deficient subspaces of $A$,
\begin{equation}
\Sigma_{z}=\text{ker}A^\dagger(z^*)=\left\{\xi_{z^*}:A^\dagger \xi_{z^*}=z^*\xi_{z^*}\right\}, \quad  \Sigma_{z^*}=\text{ker}A^\dagger(z)=\left\{\xi_{z}:A^\dagger \xi_{z}=z\xi_{z}\right\},
\end{equation}
where $z\in\mathbb{C}_{+}$ is arbitrary, but fixed, and let $m_{\pm}$ be the deficiency indices of $A$, $m_{+}=\text{dim} \Sigma_{z^*}$ and $m_{-}=\text{dim} \Sigma_{z}$. The symmetric operator $A$ has SA extensions $A_{U}=A^{\dagger}_{U}$, $A\subseteq A_{U}$ iff both its deficient subspaces $\Sigma_{z^*}$ and $\Sigma_{z}$ are isomorphic, or iff its deficiency indices  are equal, i.e. $m_{\pm}=m$. If the deficient subspaces are trivial, that is $m_{\pm}=0$, the operator $A$ is essentially SA, and its SA extension is given by its closure $\overline{A}=(A^\dagger)^\dagger$, which coincides with its adjoint, $\overline{A}=(\overline{A})^\dagger=A^\dagger$. If its deficient subspaces are nontrivial, i.e. $m_{\pm}=m\neq 0$, there exists an $m^2$-parameter family $\left\{A_{U}\right\}$ of SA extensions that is a unitary group $U(m)$. Every SA extension $A_{U}$ is determined by an isometric mapping $U: \Sigma_{z^*}\to\Sigma_{z}$ of one of the deficient subspaces onto another and is given by
\begin{equation}\begin{split}\label{a}
D_{A_{U}}&=D_{\overline{A}}+(I+U)\Sigma_{z^*}\\
&=\left\{\xi_{U}: \xi_{U}=\underline{\xi}+\xi_{z}+\hat{U}\xi_{z},\  \forall\underline{\xi}\in D_{\bar{A}},\ \forall\xi_{z}\in \Sigma_{z^*},\ U\xi_{z}\in\Sigma_{z}\right\},
\end{split}\end{equation}
where $D_{\overline{A}}$ is the domain of the closure $\overline{A}$, and
\begin{equation}\label{b}
A_{U}\xi_{U}=\overline{A}\underline{\xi}+z\xi_{z}+z^* U\xi_{z}.
\end{equation}
In other words, an  isometry $U: \Sigma_{z^*}\longrightarrow\Sigma_{z}$ that establishes an isomorphism between the deficient subspaces defines an SA extension $A_{U}$ of a symmetric operator $A$ given by \eqref{a} and \eqref{b}. If the deficient subspaces are finite-dimensional, $0<m<\infty$, then SA extensions $A_{U}$ can be specified in terms of unitary matrices $U\in U(m)$. Namely, let $\left\{e_{z,k}\right\}^{m}_{1}$ and $\left\{e_{z^*},l\right\}^{m}_{1}$ be some orthogonal bases in the respective deficient subspaces $\Sigma_{z^*}$ and $\Sigma_{z}$. Then an SA extension $A_{U}$ is given by
\begin{equation}
A_{U}:\begin{cases} D_{A_{U}}=\begin{cases} \xi_{U}: \xi_{U}=\underline{\xi}+\sum^{m}_{k=1}c_{k}e_{U,k},\ \forall \underline{\xi}\in D_{\overline{A}},\\ \forall c_{k}\in\mathbb{C}, e_{U,k}=e_{z,k}+\sum^{m}_{l=1}U_{lk}e_{z^*,l},   \end{cases},\\ A_{U}\xi_{U}=\overline{A}\underline{\xi}+\sum^{m}_{k=1}c_{k}\left(ze_{z,k}+z^*\sum^{m}_{l=1}U_{lk}e_{z^*,l}\right),\end{cases}
\end{equation}
where $U=||U_{lk}||$ is a unitary matrix.\\

\section{Some basics of rigged Hilbert space}

In order to find the full spectrum of an observable by solving a differential equation one needs to so called rigged Hilbert space \cite{Zeidler, Madrid, madrid, rigged}. Namely, if one wants to find the full spectrum of the Hamiltonian by solving the Schr\"odinger equation 
\begin{equation}\label{sche}
H\psi=E\psi
\end{equation}
then for the continuous $E$, one needs to ``look'' for solutions outside the Hilbert space $L^2(\mathbb{R})$, since the solutions in $L^2(\mathbb{R})$ exist only for discrete $E$ (point spectrum). The basic idea of the rigged Hilbert space is to find solutions to \eqref{sche} that don’t lie in $L^2(\mathbb{R})$, but rather in the adjoint of a densely defined subspace called the Schwartz space $\mathcal{S}(\mathbb{R})$. This was one constructs a Gelfand triple
\begin{equation}
\mathcal{S}(\mathbb{R})\subset L^2(\mathbb{R}) \subset\mathcal{S}^\prime(\mathbb{R}).
\end{equation}
The Schwartz space $\mathcal{S}(\mathbb{R})$ is the function space of all functions whose derivatives (multiplied by any monomial $x^\alpha$) are rapidly decreasing (often called  test functions), while $\mathcal{S}^\prime(\mathbb{R})$ is the dual of $\mathcal{S}(\mathbb{R})$, namely the space of all continuous linear functionals $\mathcal{S}^\prime(\mathbb{R}):\mathcal{S}(\mathbb{R})\longrightarrow\mathbb{C}$ (often called distributions or generalized functions).\\

Note that a rigged Hilbert space is not an extension of physics or QM, but rather the most natural mathematical structure required to study QM. Rigged Hilbert space is actually the equipping of a Hilbert space with the theory of distribution, and provides the full mathematical foundation for the so-called Dirac's bra-ket notation and well known Dirac's $\delta$-function \cite{Gieres:1999zv, Greenberger:2009zz}. In Dirac's notation the ket's $\left|\psi\right\rangle$ are elements of $\mathcal{S}(\mathbb{R})$, while the bra's $\left\langle \psi\right|$ are elements of the dual $\mathcal{S}^{\prime}(\mathbb{R})$. \\

\section{Scale symmetry in classical physics}
Here let us discuss the scale symmetry for classical systems. First, let us remind ourselves of the Noether's theorem. If the Lagrangian $L(q,\dot{q},t)$ transforms under small perturbations
\begin{equation}
t\longrightarrow t^\prime=t+\delta t, \quad q\longrightarrow q^\prime=q+\delta q
\end{equation}
so that it leaves the action $S=\int dt L(q,\dot{q}, t)$ invariant, then there exist a quantity (or many of them $\alpha=1,..., N$)
\begin{equation}\label{cq}
\mathcal {A}_\alpha:=\left(\frac{\partial L}{\partial\dot{q}}\dot{q}-L\right)T_{\alpha}-\frac{\partial L}{\partial \dot{q}}Q_{\alpha}
\end{equation}
that is conserved in time, i.e. constant of motion. Here the resulting perturbations $\delta t$ and $\delta q$ are small and can be written as
\begin{equation}\label{inf}
\delta t=\sum_{\alpha}\epsilon_\alpha T_{\alpha}, \quad \delta q=\sum_{\alpha}\epsilon_\alpha Q_{\alpha}
\end{equation}
where $\epsilon_\alpha$ are infinitesimal, $T_\alpha$ are generators of time evolution and $Q_\alpha$ are generators of generalized coordinates.\\

Now we turn to scale symmetry. If we want that the action $S$ and the equation of motions are invariant under change of scale $t\longrightarrow \lambda t$ then we can derive the following transformations for $L, q, \dot{q}, H$ and $p$
\begin{equation}\begin{split}\label{trans}
t&\longrightarrow t^\prime=\lambda t\\
L&\longrightarrow L^\prime=\lambda^{-1} L\\
q&\longrightarrow q^\prime=\lambda^{\frac{1}{2}} q\\
\dot{q}&\longrightarrow \dot{q}^\prime=\lambda^{-\frac{1}{2}} \dot{q}\\
p&\longrightarrow p^\prime= \lambda^{-\frac{1}{2}} p\\
H&\longrightarrow H^\prime=\lambda^{-1} H
\end{split}\end{equation}
Equations \eqref{trans} we call scale transformations and if the action (and equations of motion) are invariant under them, we say the system is scale invariant. If we assume that the Lagrangian is given by the difference of kinetic and potential energy, $L=T(\dot{q})-V(q)$, then the Hamiltonian is given by $H(p,q)=\left(\frac{\partial L}{\partial\dot{q}}\dot{q}-L\right)=\frac{p^2}{2m}+V(q)$. In this case, scale transformations \eqref{trans} induce a transformation for the potential energy as
\begin{equation}\label{transv}
V(q)\longrightarrow V^\prime(q)=V(\lambda^{\frac{1}{2}}q)=\lambda^{-1} V(q)
\end{equation}
Now we want to use the Noether's theorem and find the conserved quantity \eqref{cq}. In order to do so, we need to look at the infinitesimal version of \eqref{trans}. Since $\lambda$ is dimensionless we can parametrize it as
\begin{equation}
\lambda=\text{e}^\epsilon= 1+\epsilon +O(\epsilon^2)
\end{equation}
which together with \eqref{trans} and \eqref{inf} leads to 
\begin{equation}
\delta t=\epsilon t, \quad \delta q=\epsilon \frac{q}{2} \ \Longrightarrow \ \epsilon_\alpha=\epsilon, \quad T_\alpha=t, \quad Q_{\alpha}=\frac{q}{2}
\end{equation}
and the conserved quantity \eqref{cq} is\footnote{where we used $H(p,q)=\left(\frac{\partial L}{\partial\dot{q}}\dot{q}-L\right)=\frac{p^2}{2m}+V(q)$, $L=\frac{m\dot{q}^2}{2}-V(q)$ and $p=\frac{\partial L}{\partial\dot{q}}=m\dot{q}$. }
\begin{equation}\label{DD}
\mathcal {A}_\alpha=H t-\frac{1}{2}qp\equiv D.
\end{equation}
The constant of motion $D$ is called dilatation and is the generator of scale symmetry.\\

It is instructive to check that the dilatation $D$ given in \eqref{DD} is indeed the constant of motion. For that purpose we calculate $\frac{\d D}{\d t}$ using Poisson bracket formalism. Namely,
\begin{equation}\begin{split}
\frac{\d  D}{\d t}&=\frac{\partial D}{\partial t}+\left\{D,H\right\}\\
&=H-\frac{1}{4m}\left\{qp,p^2\right\}-\frac{1}{2}\left\{qp,V(q)\right\}\\
&=H-\frac{p^2}{2m}+\frac{q}{2}\frac{\partial V}{\partial q}\\
&=V(q)+\frac{q}{2}\frac{\partial V}{\partial q}.
\end{split}\end{equation}
Now, in order to have scale symmetry, not any potential energy $V(q)$ will provide the symmetry, but rather special ones that transform according to this symmetry. Remember that scale transformations \eqref{trans} induce the transformation of the potential energy \eqref{transv}, which for infinitesimals looks like
\begin{equation}
V(\lambda^{\frac{1}{2}}q)=\lambda^{-1}V(q)\ \ \Longrightarrow \  \ V(q+\frac{\epsilon}{2}q)=V(q)+\epsilon\frac{q}{2}\frac{\partial V}{\partial q}=(1-\epsilon) V(q)\ \Longrightarrow \frac{q}{2}\frac{\partial V}{\partial q}=-V(q)
\end{equation}
rendering the desired result $\frac{\d D}{\d t}=0$. In other words, systems that have potential energy $V(q)$ that satisfy the equation
\begin{equation}\label{Deq}
\frac{q}{2}\frac{\partial V}{\partial q}=-V(q)
\end{equation}
are scale invariant. Notice that the equation \eqref{Deq} actually says that the potential energy $V(q)$ has to be a solution of  the eigenvalue equation for the operator\footnote{This equation can be generalized to higher dimensions by $q\longrightarrow q^i$ and $d=q\frac{\partial}{\partial q}\longrightarrow q^i\nabla_i$.} $d=q\frac{\partial}{\partial q}$ with the eigenvalue $-2$, and the nontrivial solution\footnote{Here we have assumed that the potential energy $V(q)$ is of class $\mathcal{C}^1(\mathbb{R})$, but notice that there are also generalized functions or distributions that satisfy $V(\lambda^{\frac{1}{2}}q)=\lambda^{-1}V(q)$.} is $V(q)\propto q^{-2}$.\\

\end{document}